\documentclass[preprint,12pt]{elsarticle}

\usepackage[letterpaper,top=1.2in,bottom=1.5in,left=1.0in,right=1.0in]{geometry}%

\usepackage{amssymb,amsmath}
\usepackage[cal=rsfs]{mathalpha}
\usepackage{MnSymbol,bbding,pifont}
\usepackage{extarrows}
\usepackage{soul}
\usepackage{hyperref}
\usepackage{eso-pic}  
\usepackage{gensymb}
\hypersetup{hidelinks,
	colorlinks=true,
	allcolors=black,
	pdfstartview=Fit,
	breaklinks=true}

\usepackage{color}
\usepackage{courier}
\usepackage{booktabs}
\usepackage{makecell}
\usepackage{multirow}
\usepackage{siunitx}
\usepackage{bm}
\usepackage[figuresright]{rotating}
\usepackage[dvipsnames]{xcolor}
\usepackage[most]{tcolorbox}
\usepackage[toc,page,header]{appendix}
\usepackage{minitoc}
 
\usepackage{xcolor}

\usepackage{chngcntr} % 用于重设计数器

\usepackage[switch]{lineno}
\renewcommand\appendix{\par
    \setcounter{section}{0}
    \setcounter{subsection}{0}
    \gdef\thesection{\Alph{section}}}

{}
{}
{}
{}

\journal{AAAA}

\begin{document}
\doparttoc % Tell to minitoc to generate a toc for the parts
\faketableofcontents % Run a fake tableofcontents command for the partocs
\parttoc % Insert the document TOC

\AddToShipoutPictureFG*{%
  \AtPageLowerLeft{%
    \put(72,40){\itshape Accepted by Communications Physics--Nature}
  }%
}

\begin{linenumbers}
\begin{frontmatter}
 


% \title{Proportion-Integral-Derivative Accelerated Optimizer for Deep Learning}

\title{Transformer-based Neural Operators for 3D Wind Field Prediction over Complex Mountainous Terrain}

% \author[inst1]{Name\corref{cor1}}
% % \ead[url]{home page}
% \cortext[cor1]{Corresponding author}

\author[A,C]{Yujia Zhang\fnref{equal}}
\author[A]{Jiaxi Qi\fnref{equal}}
\author[D,D2]{Ruiyan Chen}
\author[D,D2]{Yong Liu}
\author[C]{Yuzhou Zhang}
\author[C]{Lyulin Kuang}
\author[C]{Rita Zhang}
\author[A]{Shengze Cai\corref{cor1}}\ead{shengze\_cai@zju.edu.cn}

\cortext[cor1]{Corresponding author}

\affiliation[A]{organization={College of Control Science and Engineering},%Department and Organization
            addressline={Zhejiang University}, 
            city={Hangzhou},
            country={China},
            }
\affiliation[C]{
            addressline={NVIDIA}, 
            city={Beijing},
            country={China},
            }
\affiliation[D]{
            addressline={Windey Energy Technology Group Co., Ltd.}, 
            city={Hangzhou},
            country={China},
            }
\affiliation[D2]{
            addressline={Zhejiang Key Laboratory of Offshore Wind Power Technology}, 
            city={Hangzhou},
            country={China},
            }
% \affiliation[B]{organization={State Key Laboratory of Industrial Control Technology},
%             addressline={Zhejiang University}, 
%             city={Hangzhou},
%             country={China},
%             }
% \affiliation[E]{the authors contribute equally in this paper}
\fntext[equal]{These authors contributed equally to this work.}

\begin{abstract}
Accurate prediction of three-dimensional (3D) wind fields over complex mountainous terrain is essential for renewable energy deployment and regional weather modeling. Traditional computational fluid dynamics (CFD) simulations face two fundamental bottlenecks: expert-intensive mesh generation around irregular topography, and iterative solvers that require hours to days even on high-performance clusters. Recent neural operator approaches accelerate inference, but typically fail to resolve the sharp, localized velocity gradients induced by complex terrain features. Here, we present a transformer-based dual-attention neural-operator framework for 3D wind field prediction over complex mountainous terrain, and validate its effectiveness through two instantiations on representative point-based (mesh-free) and graph-based neural-operator architectures, namely Patch-solver and Patch-GTO. Trained on a large CFD-generated dataset spanning diverse terrain geometries and inflow conditions, the framework enables rapid prediction of steady-state wind field while maintaining competitive accuracy. It also demonstrates robust zero-shot transfer to real-world mountainous sites across several diverse locations, outperforming existing neural operator baselines by 10\% in relative error. We further verify that incorporating sparse observational data (1\% spatial coverage) reduces prediction error by 16.89\% relative to the corresponding model without sparse data input and by 32.75\% relative to advanced neural operator baselines on unseen terrains. This framework establishes a generalizable computational paradigm across domains, promising to be a real-time tool for wind resource assessment over complex mountainous terrain and related atmosphere–surface interaction studies.
\end{abstract}

\end{frontmatter}


\section{Introduction}
\label{sec:intro}

Accurate prediction of three-dimensional (3D) wind fields over complex mountainous terrain is of critical importance to both renewable energy development and local meteorological applications \cite{xu2025cross}. In the context of wind energy, detailed 3D wind-flow information facilitates optimal turbine siting and accurate resource assessment, thereby directly affecting power generation efficiency and operational safety \cite{wang2022mesoscale,gao2024prediction}. Meanwhile, high-resolution wind-field forecasting is indispensable for microscale meteorological applications, such as predicting localized weather phenomena and ensuring aviation safety in complex terrains, where terrain-induced flow variability exerts substantial influence \cite{achermann2024windseer,chen2025three}. Due to the inherently unsteady and topographically modulated nature of atmospheric flows in mountainous regions, even subtle terrain features (e.g., narrow valleys or steep slopes) can induce pronounced spatial variations in wind speed and direction. This inherent complexity renders reliable 3D wind-field prediction a fundamental requirement for wind-farm planning, turbine control, hazard mitigation, and the seamless integration of wind energy into power grids \cite{zhu2021wind,yang2025wd,yang2024ultra}.

Traditionally, computational fluid dynamics (CFD) simulations and other physics-based models have served as the primary means for investigating and predicting wind fields over complex terrains. High-fidelity CFD approaches, such as large-eddy simulation (LES), are capable of resolving terrain-induced flow structures and turbulent features with high accuracy \cite{ren2018numerical}, whereas mesoscale models (e.g., WRF) are typically employed to represent larger-scale atmospheric processes. In addition, physical experiments, such as boundary-layer wind-tunnel tests, have been extensively conducted to examine the influence of terrain on wind profiles \cite{an2023experimental}. However, directly applying these methods to real-world mountainous environments remains highly challenging. CFD simulations typically involve solving flow fields over domains containing millions of mesh cells, leading to computational times ranging from several hours to days for a single case—far from feasible for iterative optimization or real-time forecasting \cite{ti2021artificial}. Even multi-scale modeling strategies, for example, nesting mesoscale models with localized LES to resolve fine-scale urban or mountainous flows, still incur considerable computational costs \cite{huang2024multi}. An inherent trade-off exists between accuracy and efficiency: coarse-grid simulations offer faster computation but fail to capture essential flow structures, whereas fine-grid simulations improve accuracy at a prohibitive computational cost \cite{kim2024enhancement}. Consequently, conventional methods are inadequate for delivering rapid and reliable 3D wind predictions over complex terrains. Recent hybrid approaches have attempted to alleviate these limitations by integrating sparse on-site measurements with CFD outputs to accelerate and correct wind resource assessments \cite{ge2025middle}. Nevertheless, a unified and computationally efficient predictive framework remains urgently needed.

In recent years, deep learning (DL) has become a promising paradigm for the rapid prediction of wind resources, and it has achieved strong performance in global weather forecasting at spatial resolutions of about $0.25\degree$ (e.g., NVIDIA Earth-2 \cite{Mardani2025}, Pangu-Weather \cite{Bi2023}, FuXi \cite{chen2023fuxi}). However, such coarse resolutions are insufficient for microscale, three-dimensional wind–field prediction over complex terrain. Operator learning offers a complementary route. Unlike conventional numerical solvers that iteratively integrate the Navier–Stokes equations on a fixed mesh, neural operators approximate the underlying solution operator of the governing partial differential equations (PDEs) by learning a nonlinear map from inputs (e.g., terrain geometry, boundary conditions) to outputs (e.g., wind velocity field), enabling resolution-robust inference and rapid generalization across diverse inputs. Early milestones include the Fourier Neural Operator (FNO) \cite{li2020fourier}, which parameterizes integral kernels in spectral space to achieve resolution-invariant inference and orders-of-magnitude speed-ups over classical CFD, and DeepONet \cite{lu2021learning}, which realizes the operator-approximation theorem in a practical branch–trunk architecture for mapping input functions to output fields. Once trained, these models achieve inference speeds several orders of magnitude faster than conventional solvers while preserving high accuracy, and have demonstrated remarkable success in turbulence modeling, climate simulation, and geophysical field prediction \cite{peng2024fourier,gao2024rapid,li2023solving,he2024geom}.

As applications move from idealized domains to real-world geometries, a central challenge is handling irregular, unstructured discretizations without sacrificing fidelity. Geometry-aware designs such as the Geometry-Informed Neural Operator (GINO) \cite{li2023geometry} encode shapes via signed-distance fields and shuttle information between graph representations and a regular latent grid, enabling spectral updates while remaining discretization-robust on complex 3-D PDEs. In parallel, Transformer-based operators address long-range coupling and multi-input fusion on irregular meshes; for example, the General Neural Operator Transformer (GNOT) \cite{hao2023gnot} introduces heterogeneous normalized attention and geometric gating to accommodate multiple fields and topologies, while recent “physics-aware tokenization” strategies, exemplified by Transolver \cite{wu2024transolver}, group points into learnable slices and attend among slice tokens to retain domain-wide coherence at near-linear cost. AeroGTO \cite{liu2025aerogto} combines local message passing from graph neural networks with global self-attention from Transformers, establishing an efficient Graph–Transformer Operator for large-scale aerodynamic predictions. Recent studies have increasingly incorporated locality, multiscale structure, and geometry awareness to alleviate the over-smoothing induced by purely global operations. MNO \cite{wang2025mno} introduces an explicit three-scale design that combines global dimension-shrinkage attention, graph attention and microscale pointwise attention. GAOT \cite{wengeometry} emphasizes multiscale and geometry-aware representation learning through graph-based encoders and decoders, while using a vision transformer as the latent space processor. Miguel et al. \cite{liu2024neural} introduce localized integral and differential kernels into FNO (global convolutions in the Fourier space) to capture local receptive fields while retaining discretization-agnostic operator learning and resolution generalization. Based on this philosophy, PCNO \cite{zeng2025point} combines integral and differential operators on point clouds to capture local effects over complex and variable geometries. Collectively, these developments mark a shift from mesh-bound, instance-wise solvers toward fast, geometry-aware operator surrogates that transfer across resolutions and unstructured grids, opening a practical path to high-fidelity predictions in fluids, climate and geoscience.

Despite these architectural advances in operator learning, existing benchmarks predominantly focus on relatively fixed geometric configurations, such as a limited set of baseline vehicle models in automotive aerodynamics (e.g., DrivAerNet \cite{elrefaie2024drivaernet} and DrivAerML \cite{ashton2024drivaerml}), simple two-dimensional airfoils (AirfRANS \cite{bonnet2022airfrans}) or cylindrical pipes \cite{li2020fourier}, or canonical flow problems with controlled geometric variations. These standardized scenarios enable systematic evaluation but do not adequately capture the challenges posed by spatially complex and highly heterogeneous terrain distributions encountered in mountainous wind field prediction. Moreover, many existing approaches \cite{gao2024prediction,achermann2024windseer,li2023geometry,nowak2024optimisation,campbell2025domain} simplify the solution domain to regular Cartesian grids, which introduces several limitations. Such uniform discretization tends to under-resolve extreme values in regions with sharp local gradients, particularly critical near ridges, valleys, and flow separation zones, and fails to represent the irregular geometric boundaries inherent in complex mountainous terrain. 

To this end, we present an end-to-end deep neural operator framework that enables rapid, high-fidelity prediction of 3D wind fields over complex mountainous terrain that would otherwise require computationally expensive CFD simulations. The framework integrates a dual-attention mechanism that simultaneously captures meter-scale terrain-induced flow features and global atmospheric patterns across the full domain, addressing the multiscale coupling inherent in mountainous wind prediction. Unlike prior approaches that define locality through graph neighborhoods \cite{wang2025mno, wengeometry} or inject local inductive bias through differential- or kernel-based operators \cite{liu2024neural, zeng2025point}, our method explicitly partitions the spatial domain into physically bounded voxels and applies learned self-attention within each voxel to capture the local flow structures induced by terrain blocking. Coupled with a separate slice-based global attention pathway, the framework further captures domain-wide flow organization, making it particularly well suited to mountainous wind prediction.

\begin{figure}[htbp!]
\centering
\includegraphics[width=\textwidth]{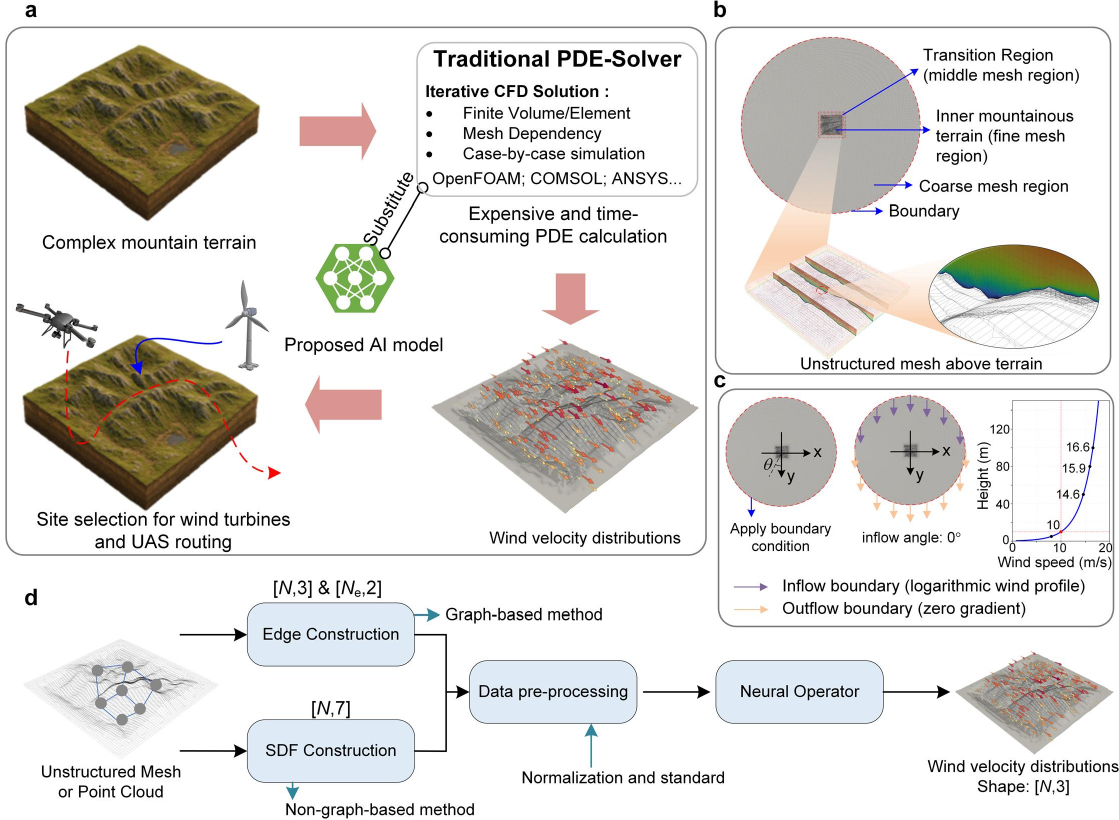}
\caption{\textbf{Illustration of the wind field dataset construction and modeling workflow.}
\textbf{a} Practical application of the proposed method, which replaces expensive PDE solvers with a mesh-agnostic neural operator to produce 3D wind velocity fields for wind turbine siting and UAS routing. \textbf{b} The entire computational domain consists of three parts: the inner fine mesh region, the transition region, and the outer coarse mesh region. The inner region, incorporating the mountainous terrain, is extracted from the circular computational domain with a size of $6000\times6000\times800~\mathrm{m}$ (width × length × altitude), serving as the core training domain for operator–based wind field prediction. \textbf{c} Boundary conditions are applied based on the inflow angle, where the windward semicircle receives a logarithmic wind profile (inflow boundary, purple arrows) with a reference velocity of 10~m/s at 10~m height, while the leeward semicircle is assigned a zero-gradient condition (outflow boundary, orange arrows). The inflow velocity profile follows a logarithmic law, as shown in the right subplot. \textbf{d} The input unstructured mesh or the point cloud is processed through two alternative pathways: Edge Construction for graph-based methods or SDF Construction for non-graph-based methods. After data pre-processing, normalization, and standardization, the neural operator learns to map the mesh representation to wind velocity distributions with shape [$N$,3], where $N$ denotes the number of mesh nodes.}
\label{fig:dataset}
\end{figure}

We validate this dual-attention principle through two instantiations on representative point-based and graph-based neural-operator architectures, resulting in Patch-solver and Patch-GTO. In particular, for Patch-solver, as shown in Fig. \ref{fig:dataset}, by representing terrain as point clouds rather than structured grids and formulating wind prediction as an operator learning problem, our approach directly maps arbitrary topographic geometries to full volumetric flow fields, eliminating the need for mesh generation and iterative PDE solvers. To validate our framework, we construct a high-fidelity CFD-based dataset spanning over 500 complex terrain regions with diverse morphologies and inlet conditions, establishing a physically consistent benchmark for operator learning in atmospheric flow prediction. Extensive comparisons against advanced neural operators reveal that our method achieves at least 10\% lower prediction errors on held-out terrains while maintaining real-time inference suitable for operational wind forecasting and renewable energy siting. In addition, the framework demonstrates three critical capabilities for operational deployment: (1) data assimilation of sparse in-situ measurements (less than 1\% spatial coverage), leading to further improvements in prediction accuracy; (2) mast-to-mast wind speed inference from annual-mean measurements at a reference mast; and (3) physically interpretable learned representations, as evidenced by entropy analysis revealing height-stratified attention patterns that mirror atmospheric boundary-layer evolution—concentrated near complex terrain where flow separation dominates, and diffuse aloft where free-stream conditions prevail. Together, these results highlight the potential of the proposed neural operator method to accelerate wind-field prediction workflows while retaining strong physical fidelity.

\section{Results}
\label{sec:res}

% \subsection{Problem definition and Proposed Method description}
% \label{sec:prob}

\subsection{Problem Setup}

Accurate prediction of wind fields over complex mountainous terrain is critical for various applications, including renewable energy siting and micro-meteorological modeling. Traditional CFD solvers, such as Fluent and OpenFOAM, can provide highly accurate predictions, but they are computationally expensive and impractical for real-time applications, especially when considering the irregularity and high computational demand of mountainous terrains. In addition, a critical bottleneck arises in wind resource evaluation workflows: converting point cloud data from terrain surveys into CFD-compatible meshes introduces expert-intensive delays that fundamentally limit the pace of resource assessment in unmapped mountainous regions.

Generally, neural operators (Non-graph-based methods) offer a transformative alternative by learning solution mappings directly from terrain geometry to flow fields, eliminating mesh generation and iterative PDE solvers once trained. Mathematically, given terrain geometry as point cloud $\mathbf{T}\in \mathbb{R}^{N_\text{s} \times C_\text{in}}$ and query locations $\mathbf{P} \in \mathbb{R}^{N \times 3}$ in the 3D atmospheric domain, the problem is to learn a neural operator $\Phi$ that predicts the velocity field:
\begin{equation}
\begin{aligned}
\hat{\mathbf{W}} \in \mathbb{R}^{N \times C_\text{in}}= \Phi([\mathbf{T},\mathbf{P}])
\label{eq1: problem}
\end{aligned}
\end{equation}
where $\mathbf{P}\in \mathbb{R}^{N \times C_\text{in}}$ represents the coordinates of points above the given terrain ($\mathbf{T}\in \mathbb{R}^{N_\text{s} \times C_\text{in}}$, where $N_\text{s}$ is the number of points of the given terrain); $\hat{\mathbf{W}} = [\hat{u}, \hat{v}, \hat{w}]$ represents the predicted velocity components at each query point.

However, applying neural operators to mountainous wind prediction poses a fundamental challenge: terrain-induced flows exhibit multi-scale coupling where global pressure-velocity interactions spanning kilometers determine overall circulation, while meter-scale terrain features (ridges, valleys, steep slopes) induce localized flow separation and sharp velocity gradients that must be simultaneously resolved. Accordingly, methods dominated by global aggregation \cite{wu2024transolver} may smooth out local structures, whereas methods with a strong local inductive bias \cite{li2023geometry} may insufficiently capture long-range dependencies.

\subsection{Transformer-based neural operators with dual-attention architecture}

We resolve this multi-scale challenge through a principled decomposition of the solution operator into complementary spatial pathways. Our physics-motivated dual-attention architecture integrates two attention mechanisms that explicitly encode the scale separation inherent in terrain-flow interactions:

\textbf{(i)} Local sectional attention partitions the atmospheric domain into spatially contiguous voxels and computes self-attention exclusively within each voxel neighborhood. This voxel-based processing is designed to preserve sharp velocity gradients and capture terrain-confined flow structures—separation bubbles, recirculation zones, and boundary-layer detachment—that dominate near complex topography.

\textbf{(ii)} Global slice attention compresses the entire point cloud into a small set of learnable slice tokens via soft assignment, then applies self-attention among these tokens to capture domain-wide pressure-velocity coupling and large-scale circulation patterns. This compressed representation maintains atmospheric coherence while remaining computationally tractable.

\textbf{(iii)} Learnable fusion combines the two pathways through a globally shared, trainable scalar gate, which balances the overall contributions of local sectional attention and global slice attention across the entire domain. The gate is optimized during training to find the best trade-off between terrain-confined local features and domain-wide flow patterns.

Critically, both attention pathways are applied across the entire 3D computational domain rather than being restricted to specific regions. However, their effective contributions are spatially heterogeneous. We validate this dual-attention principle by integrating it into two representative neural operator frameworks, Transolver \cite{wu2024transolver} and AeroGTO \cite{liu2025aerogto}, and develop two enhanced variants, Patch-solver and Patch-GTO, to deliver second-level inference latency and high-accuracy wind field predictions. A detailed description of the developed neural operators can be found in Section \ref{sec:method}.

\subsection{Dataset}
\label{sec:data}

We construct a large-scale steady-state wind field dataset from high-fidelity CFD simulations over diverse mountainous terrains based on Reynolds-averaged Navier-Stokes (RANS) simulations to train and evaluate the proposed neural operators. The dataset setup is illustrated in Fig.~\ref{fig:dataset}b-d, and the simulation details are summarized in Appendix.~\ref{OPENfoam}. The computational domain adopts a cylindrical geometry with a circular cross-section (Fig.~\ref{fig:dataset}b), consisting of three nested regions: (i) the inner region containing the actual terrain with fine mesh resolution, (ii) the transition region with gradually coarsening mesh, and (iii) the outer region with coarse mesh extending to the far-field boundaries. From this full domain, we extract a core cubic region of $6000\times6000\times800~\mathrm{m}$ (width $\times$ length $\times$ altitude) centered on the terrain, which serves as the training and inference domain for all neural operators. Boundary conditions vary with inflow angle $\theta$ to simulate different wind directions (Fig.~\ref{fig:dataset}c). The windward semicircle receives a logarithmic wind profile following atmospheric boundary layer theory~\cite{richards1993appropriate}. The leeward semicircle employs a zero-gradient outflow condition ($\partial \mathbf{U}/\partial n = 0$, where $n$ denotes the outward normal direction to the boundary) to allow natural flow development.

The dataset comprises 45 distinct terrain geometries sampled from real mountainous regions using NASA's Shuttle Radar Topography Mission (SRTM) digital elevation model (DEM) at 30~m spatial resolution \cite{farr2007shuttle}, each simulated under up to 16 uniformly spaced inflow angles ($\theta=0^\circ, 22.5^\circ, 45^\circ, \dots, 337.5^\circ$), yielding 467 terrain-angle combinations. Each wind field sample in the dataset is generated using OpenFOAM simulations, requiring approximately 1 hour of wall-clock time on a single core of an AMD EPYC 7543 32-Core Processor.

To standardize the input representation and enhance learning efficiency, we apply a coordinate transformation to align all samples to a canonical $0^\circ$ inflow direction. Specifically, for each sample with an original inflow angle $\theta$, we rotate both the terrain coordinates and the velocity field by $-\theta$ about the $z$-axis, ensuring that the dominant wind direction consistently points along the negative $y$-axis across all samples. This alignment procedure transforms the 467 terrain-angle combinations into 467 distinct terrain-flow configuration patterns, each representing a unique interaction between topographic features and atmospheric flow under standardized boundary conditions. Each sample contains approximately 360,000 spatial points with associated three-dimensional velocity fields $(u,v,w)$. Since subsequent neural operator networks employ different architectures, two distinct input representations (Fig.~\ref{fig:dataset}d) are constructed in this paper. For graph-based methods, we explicitly construct edge connectivity by combining the original mesh edges with additional edges established via k-nearest neighbor (KNN) search \cite{liu2025aerogto}, forming an edge set $\mathbf{E}\in\mathbb{Z}^{[N_\text{e},2]}$. For non-graph-based methods, we employ Signed Distance Fields (SDF) \cite{wu2024transolver} to implicitly represent geometric relationships through distance functions and surface normals. All coordinates are normalized to $[0,1]$, and velocity components are standardized to zero mean and unit variance across the training set. Detailed specifications of the computational domain design, mesh statistics, and data pre-processing procedures are provided in Appendix~\ref{Detailed dataset description}.

The ratio between training and testing sets is set to 8:2. To validate the generalization performance of the trained models, an additional zero-shot evaluation dataset is constructed, comprising four geographically distinct mountainous sites that are spatially separated from the training regions and exhibit diverse topographic characteristics. Multiple wind inlet directions are simulated for each unseen terrain to capture various wind scenarios. After applying the same coordinate transformation and preprocessing pipeline, the zero-shot dataset contains 63 test cases in total, enabling a comprehensive assessment of model robustness and transferability across different terrain morphologies.

% \subsection{Experiment setup}
In the following sections, we systematically evaluate the proposed neural operator framework across five critical dimensions that collectively establish its practical viability for operational wind resource assessment.
\begin{itemize}
    \item We benchmark prediction accuracy against previous advanced neural operator baselines on the test set, analyzing both error distributions and average performance metrics across velocity components to quantify the improvements. 
    \item We examine how prediction performance varies with terrain complexity and altitude.
    \item We assess the zero-shot generalization capability at geographically distinct mountainous sites and evaluate prediction performance across multiple inflow direction conditions to examine directional sensitivity.
    \item We evaluate the framework's capacity to assimilate sparse in-situ measurements, simulating realistic wind monitoring scenarios where limited observational data (0.1\%-1\% spatial points measured) is available.
    \item We demonstrate the practical applicability of the framework through a real-world mast-to-mast wind speed inference task, validating its capacity to capture terrain-modulated transfer relationships between meteorological masts at a complex mountainous site.
\end{itemize}

\subsection{General results for wind field prediction}

\begin{figure}[htbp!]
\centering
\includegraphics[width=\textwidth]{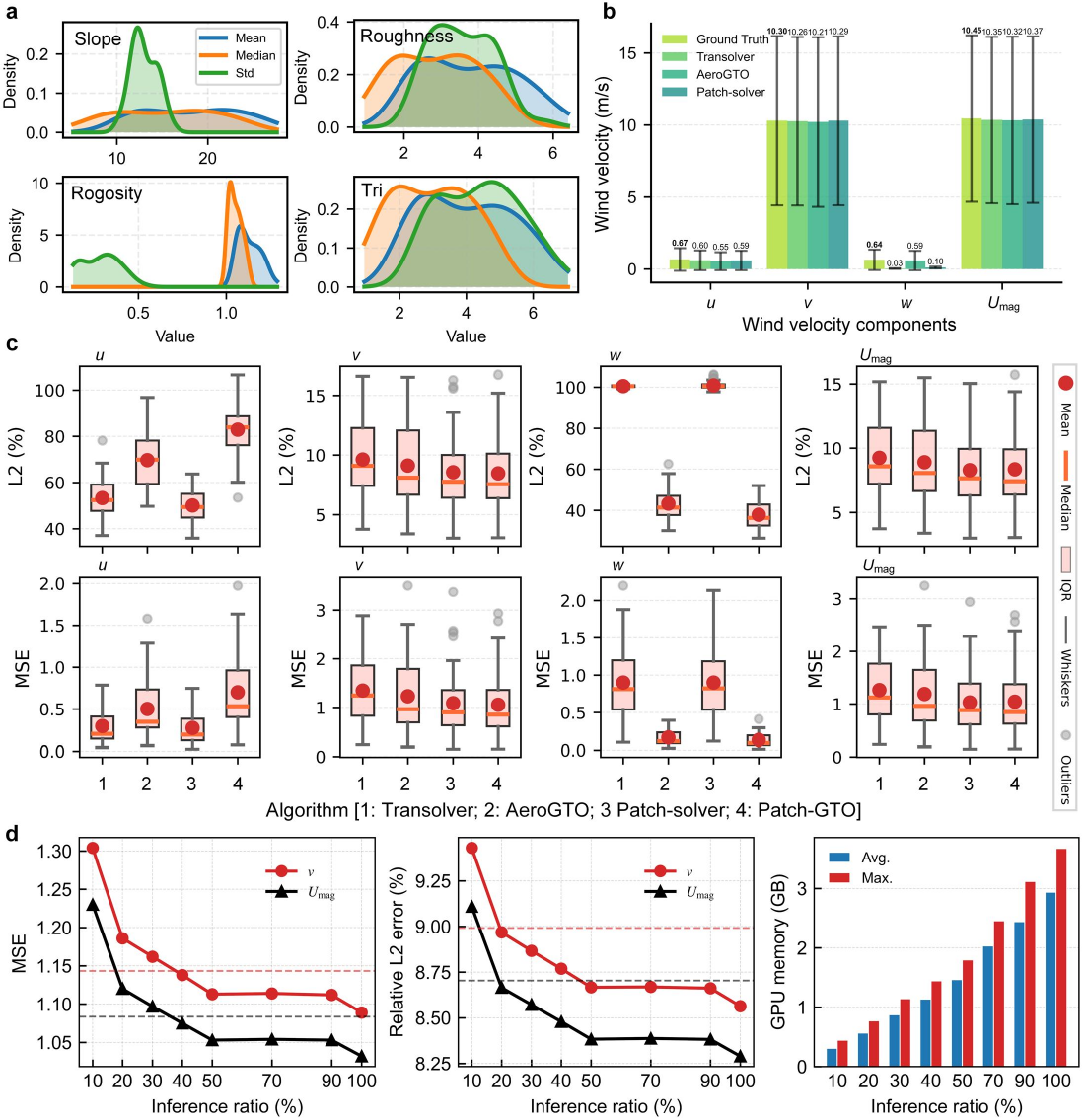}
\caption{\textbf{General wind-field prediction results across test dataset.} \textbf{a} Kernel–density distributions of four terrain descriptors—Slope, Roughness, Rugosity, and TRI—computed over the dataset (see Appendix~\ref{terrain attr}). For each descriptor, we plot the distribution of per-case mean (blue), median (orange), and standard deviation (green), delineating the range of terrain conditions represented in training and testing. \textbf{b} Dataset-averaged wind velocities for the three velocity components $(u,v,w)$ and the speed magnitude $U_{\text{mag}}$ from ground truth and four models (Transolver, AeroGTO, Patch-solver, Patch-GTO). Bars denote the mean over the entire test set; error bars indicate variability across cases. \textbf{c} Box-and-whisker summaries of various evaluation metrics (relative L2 (\%) (top row) and MSE (bottom row) for each velocity component ($u$, $v$, $w$, $U_{\text{mag}}$) across the test set. The algorithms 1–4 in the $x$-axis are Transolver, AeroGTO, Patch-solver, and Patch-GTO, respectively. Boxes show the interquartile range (IQR), central orange lines mark the median, red circles indicate the mean, whiskers span the non-outlier range, and grey dots are outliers. \textbf{d} Zero-shot resolution-robustness evaluation. The first two panels show how the MSE and relative L2 error of $u$ and $U_{\text{mag}}$ vary with the inference ratio. The dashed horizontal lines indicate the 5\% error increase relative to the corresponding error at the 100\% inference ratio. The third panel reports the average and maximum GPU memory consumption under different inference ratios.}\label{fig:result1}
\end{figure}

In this section, we report test-set errors for velocity components $(u,v,w)$ and speed magnitude $U_{\text{mag}}$ using various metrics (MSE, relative L2, and MAE), as illustrated in Table~\ref{tab:result-1}. Detailed descriptions of these metrics are provided in Eqs.~\ref{eq1: l2}-\ref{eq1: mae}. The evaluated terrain cases in the dataset span the feature distributions shown in Fig.~\ref{fig:result1}, encompassing slope, roughness, rugosity, and TRI (Terrain Ruggedness Index), with terrain descriptors computed following the procedures detailed in the Appendix.~\ref{terrain attr}. To contextualize the operating regime, we present the mean, median, and standard deviation (std) of each descriptor alongside the density distributions in Fig.~\ref{fig:result1}a. Specifically, the roughness parameter exhibits mean values concentrated between 2 and 5 with a std of around 3.5, the slope varies with means between $10\degree$ and $25\degree$ (std around 12), the TRI ranges from 2 to 6.5 in mean (std around 4), and rugosity centers around 1.0 to 1.5 (std around 0.3). These distributional characteristics delineate the terrain feature space represented in the training data, establishing the domain over which the models are expected to generalize reliably. Predictions on test cases falling within these established ranges maintain high fidelity. In contrast, inference on terrain features substantially outside this distribution may exhibit degraded accuracy due to the extrapolative nature of such queries.

% 为了便于后续的attention计算，我们在

\begin{table}[!htbp]
  \centering
  \caption{Performance comparison on the wind velocity prediction task with various algorithms}
  \label{tab:result-1}
  \setlength{\tabcolsep}{3.5pt}     % 列间距(默认 6pt)
  \renewcommand{\arraystretch}{1} % 行高(默认 1)

  \begin{tabular}{l*{12}{c}}
    \toprule
    \multirow{3}{*}{\textbf{Model}} &
      \multicolumn{12}{c}{\textbf{Metrics}} \\
    \cmidrule(lr){2-13}
      & \multicolumn{4}{c}{\textbf{MSE}}
      & \multicolumn{4}{c}{\textbf{L2} ($\%$)}
      & \multicolumn{4}{c}{\textbf{MAE}} \\
    \cmidrule(lr){2-5}\cmidrule(lr){6-9}\cmidrule(lr){10-13}
      & $u$ & $v$ & $w$ & {$U_{\text{mag}}$}
      & $u$ & $v$ & $w$ & {$U_{\text{mag}}$}
      & $u$ & $v$ & $w$ & {$U_{\text{mag}}$} \\
    \midrule
    Transolver & 0.299   & 1.348 & 0.901 & 1.264 & 53.283 & 9.626 & 100.526 & 9.257 & 0.315  &  0.706& 0.637 & 0.692  \\
    AeroGTO      & 0.502       &  1.231     &  0.169     &   1.191     & 69.709     & 9.122   &  43.290     &  8.912      &    0.418  &  0.682     &    0.268   &   0.676     \\
    GINO         &    0.841   &   20.728    &  0.654     &   19.576    &    91.451   &   38.558    &     86.778  &    37.168   &    0.582    &   3.297   &    0.532   &   3.225    \\
    GNOT         &   0.795    &   3.075    &    0.554   &    2.757   &   88.322    &  14.614     &  80.813     &    13.758   &      0.547 &     1.072  &    0.486   &  1.039     \\
    Geo-FNO        &  0.937 &    4.458   &    0.604   &  3.798     &   98.056     &   17.663   &   85.900     &   16.252    &  0.608      &    1.337  &  0.529     &    1.267   \\
    \midrule
    Patch‐solver & 0.275  & 1.089  & 0.900  & 1.032  & 50.100 & 8.564 & 100.902 & 8.290 & 0.309 & 0.642 & 0.647 & 0.635    \\
    Patch‐GTO    &  0.701     &  1.057     &    0.136   &    1.046   &   82.887    &   8.465    &   37.921  &  8.361     &  0.512     &    0.627   &    0.231   &    0.630   \\
    \bottomrule
    
  \end{tabular}
\end{table}

As shown in Table~\ref{tab:result-1}, our proposed PDE solver (Patch-solver and Patch-GTO) with dual-attention consistently outperforms established baselines across nearly all metrics. For wind speed magnitude $U_{\text{mag}}$, Patch-solver achieves the best overall performance with MSE of 1.032 and relative L2 error of 8.290\%, representing a 13.4\% reduction in MSE and 7.0\% reduction in L2 error compared to the strongest baseline, AeroGTO (MSE 1.191, L2 8.912\%). When compared to the pure Transformer-based baseline Transolver, Patch-solver reduces MSE by 18.4\% (from 1.264 to 1.032) and L2 error by 10.4\% (from 9.257\% to 8.290\%). Similarly, incorporating the dual-attention mechanism into the graph-operator framework yields comparable gains: Patch-GTO improves upon its base architecture AeroGTO by 12.2\% in MSE (from 1.191 to 1.046) and 6.2\% in L2 error (from 8.912\% to 8.361\%). These consistent improvements across both architectural paradigms directly demonstrate the effectiveness of coupling local sectional attention with global attention mechanisms for complex terrain wind field prediction. Given the cubic dependence of turbine power on wind speed \citep{veers2019grand}, a 10\% error in wind-speed prediction propagates to approximately 33\% deviation in power output estimates, therefore representing a practically meaningful improvement for wind-resource assessment and turbine siting.

Due to the coordinate transformation applied during preprocessing, the primary wind direction aligns with the $v$ axis. Both proposed models (Patch-solver and Patch-GTO) also demonstrate strong predictive capabilities, with Patch-solver achieving an average MSE of 1.089, representing a 19.2\% improvement in MSE over Transolver (MSE 1.348). This superior performance along the primary flow direction reflects the dual-attention mechanism's effectiveness in capturing terrain-modulated streamwise evolution. 

% For the $u$ component, which characterizes lateral deflections arising primarily from terrain blocking and channeling effects, Patch-solver again exhibits the strongest performance across all metrics (MSE 0.275, L2 50.10\%, MAE 0.309), outperforming Transolver by 8.0\% in MSE (from 0.299 to 0.275) and 6.0\% in L2 error (from 53.283\% to 50.10\%). These cross-flow velocities are particularly sensitive to local geometric constraints, as terrain obstacles redirect the incoming flow laterally through separation, recirculation, and sidewall boundary layers. 

However, a striking characteristic evident across all models is the dramatically higher prediction error in the vertical component $w$ and cross-flow velocities $u$ compared to the horizontal primary components $v$, as described in Table~\ref{tab:result-1}. This disparity stems from the fundamentally different physical mechanisms governing these components: the $v$ component inherits the relatively continuous characteristics of the incoming freestream flow and exhibits gradual spatial variations as it adapts to terrain-induced acceleration and deceleration, while the $w$ and $u$ components represent terrain-generated disturbances that exhibit highly irregular and spatially localized patterns. These secondary components emerge through highly nonlinear mechanisms caused by topographic blocking, producing spatially chaotic and localized distribution patterns that are inherently difficult to predict. The introduction of dual-attention mechanisms enables Patch-solver to reduce prediction errors in the $u$ direction relative to the best baseline (8.0\% MSE reduction from 0.299 to 0.275), while Patch-GTO demonstrates improved accuracy in capturing the $w$ component dynamics (19.5\% MSE reduction from 0.169 to 0.136). However, the absolute error magnitudes for these terrain-induced components remain substantial across all evaluated models.

To contextualize these error metrics, we visualize the mean values and statistical distributions of predictions versus ground truth across the entire test dataset in Fig.~\ref{fig:result1}b. Note that, due to the coordinate transformation applied during preprocessing, all samples are rotated such that the dominant inflow direction is aligned with the $v$-axis. As a result, the $u$ and $w$ components mainly represent terrain-induced lateral and vertical deflections rather than the primary streamwise flow. Consequently, the dataset-averaged wind velocity in the primary $v$ direction is 10.30~m/s, whereas the mean velocities in the $u$ and $w$ directions are only about 0.6~m/s, corresponding to roughly 5\% of the streamwise magnitude. Under such conditions, even minor absolute prediction errors (e.g., 0.3 m/s) translate into disproportionately large relative error percentages for the secondary components. Nevertheless, as evident from Fig.~\ref{fig:result1}b, the absolute discrepancies between the predicted and ground-truth distributions for these components remain small. In addition, using a common first-order estimate for yaw-related power reduction \cite{liew2020analytical}, $P(\gamma)/P(0)\approx \cos^3(\gamma)$, where $P(\gamma)$ denotes the turbine power under a flow-deflection angle $\gamma$ and $P(0)$ denotes the reference power under aligned inflow, the dataset-averaged values imply only a small reduction of approximately 0.5\% when $\gamma=\text{arctan}(0.67/10.30)=3.71^\circ$. Therefore, in the present dataset, the primary streamwise flow remains the most direct contributor to turbine power production, whereas the influence of $u$ and $w$ is more indirect. Thus, for wind resource assessment, the high prediction accuracy of the primary streamwise flow component $v$ provides a reliable basis for evaluating overall energy-yield-related performance.

% sing a common first-order estimate for yaw-related power reduction, $P(\gamma)/P(0)\approx \cos^3(\gamma)$, where $P(\gamma)$ denotes the turbine power under a flow-deflection angle $\gamma$ and $P(0)$ denotes the reference power under aligned inflow, the dataset-averaged values in our test set imply only a small reduction of approximately 0.5\% if only $u$ is considered, or about 1.0\% if both $u$ and $w$ are combined into an equivalent flow-deflection angle. Therefore, in the present dataset, the primary streamwise flow remains the most direct contributor to turbine power production, whereas the influence of $u$ and $w$ is more indirect.
% Nevertheless, from a wind resource assessment perspective, these cross-flow and vertical velocity components exert limited influence on practical energy yield estimation, as wind turbine power generation is primarily governed by the magnitude and consistency of the streamwise wind velocity-precisely the component where our proposed models demonstrate superior predictive accuracy.

Fig.~\ref{fig:result1}c summarizes case-wise error distributions on the test set. For each component ($u$, $v$, $w$, $U_{\text{mag}}$), we report the relative L2 (\%) and MSE across all test cases using box-and-whisker plots (mean as red dot, median as orange line, boxes for IQR, whiskers for the non-outlier range). As depicted in Fig.~\ref{fig:result1}c, Patch-solver shifts both the median and mean errors downward for the primary wind component $v$ and for wind magnitude $U_{\text{mag}}$, while also narrowing the IQR—indicating not only higher accuracy but also improved robustness across terrains. Besides, Patch-GTO most benefits the vertical component $w$, lowering central errors and trimming long tails relative to AeroGTO. Overall, the proposed dual-attention PDE solver reduces both bias (mean/median) and dispersion (IQR/whiskers) compared with the corresponding baselines, with the largest gains observed for the practically most relevant quantities, $v$ and $U_{\text{mag}}$. 

Fig.~\ref{fig:result1}d further examines the resolution robustness of the proposed framework through a zero-shot evaluation, with detailed descriptions provided in Appendix.~\ref{appen: Zero-shot super-resolution analysis}. The model is trained using the full-resolution point clouds (100\% inference ratio) and is then directly evaluated under different inference ratios without retraining. As shown in Fig.~\ref{fig:result1}d, both the MSE and relative L2 errors increase gradually as the inference ratio decreases. Nevertheless, the degradation remains relatively mild over a wide range of ratios. In particular, when the inference ratio is reduced from 100\% to 50\%, the error curves remain nearly flat, indicating that the model retains almost the same predictive accuracy even when only half of the query points are processed in a single forward pass. A more noticeable increase appears only when the ratio drops below approximately 30\%. For example, at the extreme case of 10\%, the MSE and relative L2 error of $U_{\text{mag}}$ increase by about 19\% and 9.9\% relative to the full-resolution setting. Besides, the computational cost decreases significantly with the inference ratio. Specifically, the maximum GPU memory consumption drops from approximately 3.7~GB at the full resolution to about 0.45~GB at the 10\% setting. These results demonstrate that the proposed neural-operator framework maintains strong predictive stability across varying point densities while offering flexible control over inference cost, which is particularly valuable for large-scale terrain simulations and resource-constrained deployment scenarios.

\subsection{Case study and visualization}

\begin{figure}[htbp!]
\centering
\includegraphics[width=\textwidth]{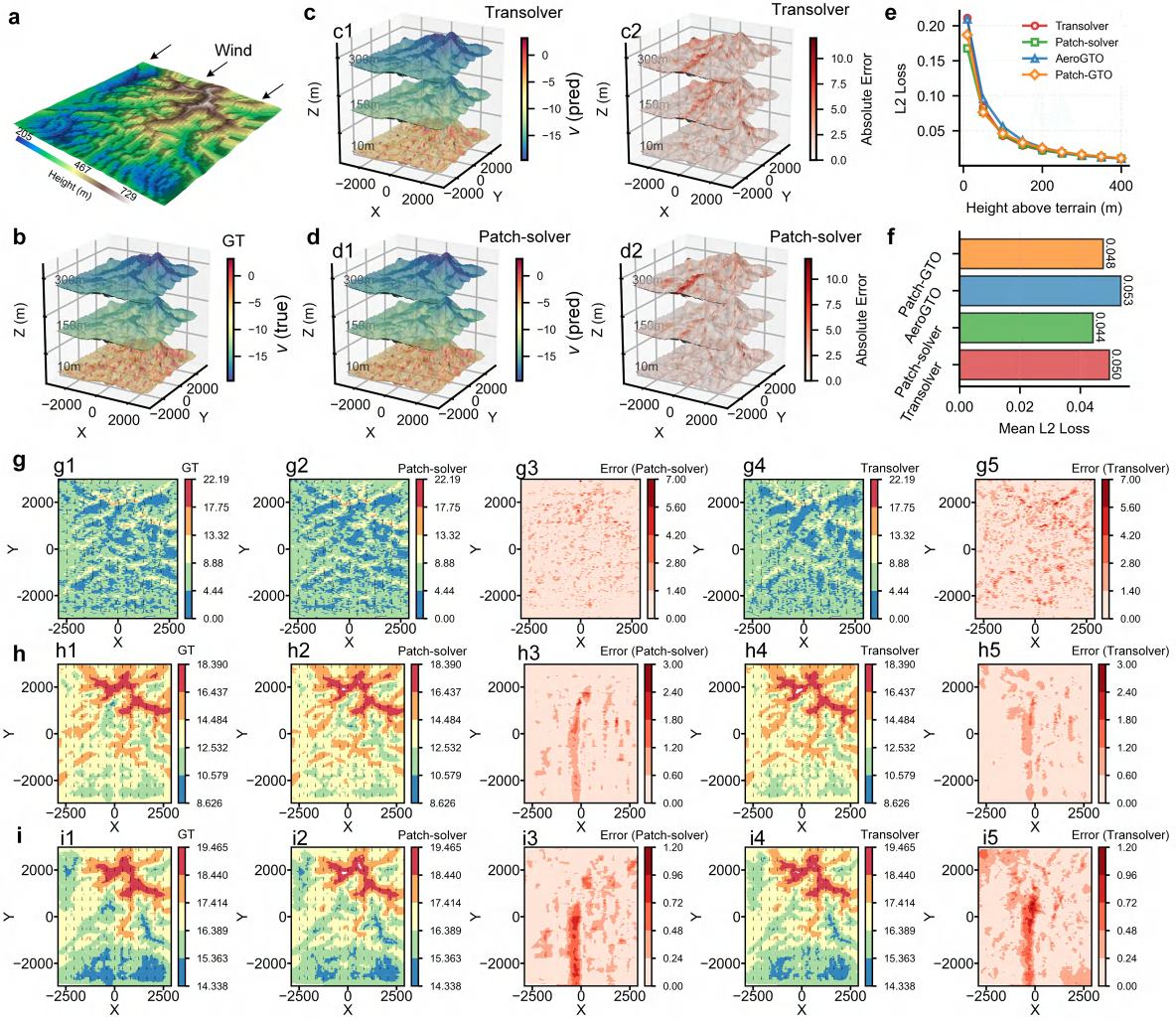}
\caption{\textbf{Altitude–resolved qualitative comparison and error analysis on a relatively flat case.}
\textbf{a} Schematic diagram of mountain terrain and wind direction. \textbf{b} Ground–truth wind–velocity component at primary direction ($v$) sampled on multiple (10m, 150m and 300m) altitude planes. \textbf{c} Prediction results (\textbf{c1}) and absolute error (\textbf{c2}) of wind-velocity component $v$ obtained by Transolver at the corresponding altitude planes in 3D viewpoint. \textbf{d} Prediction results (\textbf{d1}) and absolute error (\textbf{d2}) of wind-velocity component $v$ obtained by Patch-solver at the corresponding altitude planes in 3D viewpoint.
\textbf{e} Altitude-dependent relative L2 error averaged across the scene for all methods (Transolver, Patch-solver, AeroGTO, Patch-GTO). \textbf{f} Mean L2 error aggregated over altitude layers in \textbf{e}. \textbf{g–i}, Plan-view maps at three representative altitude planes (10m, 150m, 300m, respectively) of wind-speed magnitude $U_\text{mag}$. Columns show, from left to right: ground truth (\textbf{g1/h1/i1}), Predictions of Patch-solver (\textbf{g2/h2/i2}), Pix-level absolute error of Patch-solver (\textbf{g3/h3/i3}), Predictions of Transolver (\textbf{g4/h4/i4}), and Pix-level absolute error of Transolver (\textbf{g5/h5/i5}). Arrows denote horizontal velocity vectors. }
\label{fig:case1}
\end{figure}

\begin{figure}[htbp!]
\centering
\includegraphics[width=\textwidth]{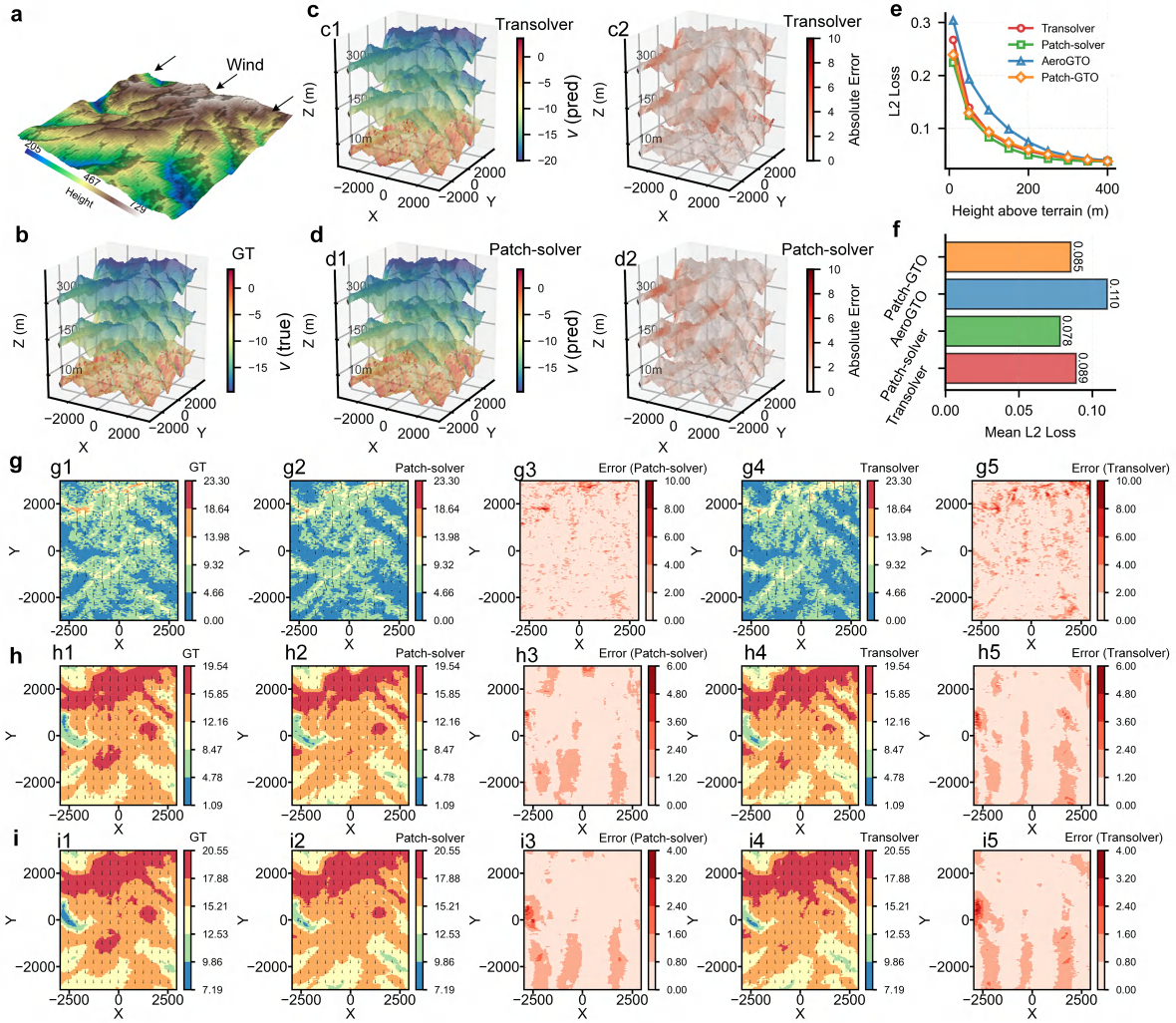}
\caption{\textbf{Altitude–resolved qualitative comparison and error analysis on a relatively complex case.}
\textbf{a} Schematic diagram of mountain terrain and wind direction. \textbf{b} Ground–truth wind–velocity component at primary direction ($v$) sampled on multiple (10m, 150m and 300m) altitude planes. \textbf{c} Prediction results (\textbf{c1}) and absolute error (\textbf{c2}) of wind-velocity component $v$ obtained by Transolver at the corresponding altitude planes in 3D viewpoint. \textbf{d} Prediction results (\textbf{d1}) and absolute error (\textbf{d2}) of wind-velocity component $v$ obtained by Patch-solver at the corresponding altitude planes in 3D viewpoint. \textbf{e} Altitude-dependent relative L2 error averaged across the scene for all methods (Transolver, Patch-solver, AeroGTO, Patch-GTO). \textbf{f} Mean L2 error aggregated over altitude layers in \textbf{e}. \textbf{g–i}, Plan-view maps at three representative altitude planes (10m, 150m, 300m, respectively) of wind-speed magnitude $U_\text{mag}$. Columns show, from left to right: ground truth (\textbf{g1/h1/i1}), Predictions of Patch-solver (\textbf{g2/h2/i2}), Pix-level absolute error of Patch-solver (\textbf{g3/h3/i3}), Predictions of Transolver (\textbf{g4/h4/i4}), and Pix-level absolute error of Transolver (\textbf{g5/h5/i5}). Arrows denote horizontal velocity vectors. }
\label{fig:case2}
\end{figure}

To comprehensively assess the predicted performance characteristics of our proposed models, we visualize wind field predictions at multiple elevations above the terrain surface through two representative test cases that span different degrees of topographic complexity, as depicted in Fig. \ref{fig:case1} and \ref{fig:case2}. For each case, we extract horizontal cross-sections at three strategically selected altitudes—10~m, 150~m, and 300~m above the terrain surface. Here, the values at these specified altitudes are extracted from the nearest grid points at the corresponding heights above the terrain surface. These elevations fall within the rotor-swept zone of wind turbines sited in mountainous terrain and thus provide the wind-velocity distribution at application-relevant heights.

The first test case, shown in Fig. \ref{fig:case1}a, features gently rolling topography with modest elevation variations, providing a scenario where terrain-induced flow perturbations are relatively mild. The ground-truth velocity field (Fig. \ref{fig:case1}b) reveals that even over this gentle terrain, the primary wind component $v$ exhibits notable spatial heterogeneity concentrated in the near-surface layer, with flow patterns progressively homogenizing at higher altitudes as the terrain influence diminishes. Three-dimensional visualizations of the predicted $v$ component and corresponding absolute errors are presented for both the Transolver and the Patch-solver in Fig. \ref{fig:case1}c-d. Visualization results reveal that the Patch-solver captures the vertical stratification of wind velocity with higher fidelity, particularly in the near-terrain region where velocity gradients are steepest. The absolute error distributions demonstrate that the Patch-solver exhibits consistently lower error magnitudes across all altitude planes, with particularly pronounced improvements at the 10~m level, where local terrain effects dominate. 

Quantitative altitude-resolved error analysis in Fig. \ref{fig:case1}e shows that all four evaluated models follow a consistent trend: the relative L2 error decreases with increasing height above the terrain. Throughout various altitudes, Patch-solver and Patch-GTO maintain consistent performance advantages over their respective baseline architectures, with the gap being most pronounced in the challenging near-surface regime. Furthermore, the altitude-integrated mean L2 error in Fig. \ref{fig:case1}f shows that Patch-solver achieves the lowest overall error of 0.044, followed closely by Patch-GTO at 0.048, while both substantially outperform Transolver (0.050) and AeroGTO (0.053), corresponding to relative error reductions of 12\% and 9.4\%, respectively.

Plan-view diagram (Fig. \ref{fig:case1}g-i) of wind speed magnitude $U_\text{mag}$ and wind direction (denoted by arrows in the figure) at three representative altitudes above the terrain provides a precise comparison with predictions and ground-truth results. At 10~m elevation (Fig. \ref{fig:case1}g), the ground truth exhibits a heterogeneous velocity field ranging from approximately 0 to 22 m/s, with lower velocities concentrated in topographic depressions and flow acceleration occurring over ridges. Patch-solver predictions (Fig. \ref{fig:case1}g2) closely reproduce these spatial patterns, with pixel-wise absolute errors (Fig. \ref{fig:case1}g3) remaining predominantly below 5.6 m/s and exhibiting a relatively uniform spatial distribution. In contrast, Transolver predictions (Fig. \ref{fig:case1}g4) show more pronounced local discrepancies, particularly in regions of flow acceleration, resulting in error magnitudes (Fig. \ref{fig:case1}g5) that frequently exceed 5.6 m/s with more spatially concentrated error hot spots. Moving to 150~m altitude (Fig. \ref{fig:case1}h), where wind speeds range from 8.6 to 18.4 m/s, both models perform substantially better as the terrain influence weakens. Patch-solver errors remain below 2.40 m/s across most of the domain, while Transolver errors occasionally reach 3.00 m/s in localized regions. By 300~m altitude (Fig. \ref{fig:case1}i), with speeds spanning 14.3 to 19.5 m/s, prediction accuracy improves further for both methods. At this altitude, local terrain effects are negligible, leading to essentially identical prediction errors for Transolver and Patch-solver, with the absolute error in most areas below 1~m/s.

% \subsection{Forecasting results at various heights: Case 2}
% \subsection{Case study 2: A relatively complex mountainous terrain}

The second visualization case (Fig. \ref{fig:case2}a) presents a considerably more challenging evaluation by introducing significantly steeper elevation gradients, sharp ridgelines, and deeply incised valleys. This complex mountainous terrain provides a rigorous test of model robustness under conditions representative of actual wind farm siting locations in mountainous regions. The ground-truth $v$ component (Fig. \ref{fig:case2}b) exhibits highly irregular spatial patterns with dramatic velocity variations, such as several regions with pronounced acceleration over exposed windward slopes and ridge crests. These flow characteristics reflect the fundamentally different physics at play: whereas the gentle terrain primarily induces smooth flow deformation, the steep topography triggers flow separation, recirculation, and intermittent reattachment that generate highly localized, nonlinear velocity structures. 

Comparing the three-dimensional prediction visualizations in Fig.\ref{fig:case2}c-d reveals that Patch-solver predictions capture the complex flow structure with notably better spatial coherence, maintaining sharp gradients near topographic discontinuities while avoiding spurious oscillations. The corresponding absolute error fields show that Transolver struggles particularly in regions of strong flow curvature, such as separation zones downstream of steep ridges and re-acceleration regions in valley constrictions, where errors reach their maximum values (Fig.\ref{fig:case2}c2). The dual-attention mechanism proves especially beneficial here: by combining global physics attention to maintain large-scale pressure-velocity coupling with local sectional attention to resolve terrain-confined eddies. Patch-solver achieves substantially reduced error magnitudes throughout the computational domain, especially in the altitude plane below 150~m. Around 300~m above the terrain, the terrain influence is limited, and the predictions of Patch-solver and Transolver are nearly indistinguishable.

The altitude-dependent error (Fig.\ref{fig:case2}e) trends for this complex case reveal elevated absolute error levels compared to the flat terrain, with near-surface relative L2 errors spanning 0.22-0.30 across models, approximately 30\% higher than the previous case. This elevated error floor reflects the fundamentally more challenging prediction problem posed by abrupt terrain features that induce three-dimensional flow separation and complex vortex structures. As height increases, the errors decrease toward the levels observed in Case~1, and by 200~m the relative errors are approximately 0.03 to 0.06 for all models. The relative performance ranking remains consistent, with Patch-solver and Patch-GTO continuing to outperform their baseline counterparts across all elevations. The altitude-averaged mean L2 error (Fig. \ref{fig:case2}f) quantifies the overall challenge escalation: Patch-solver achieves 0.078 (compared to 0.044 in the flat case), while Transolver degrades to 0.089, which corresponds to an approximate 12\% reduction for Patch-solver relative to Transolver. These results indicate that the introduction of dual attention architecture remains effective even in complex terrain.

Figs. \ref{fig:case2}g-i visualize the velocity magnitude slices at different altitudes above the terrain. At 10~m altitude, the ground truth exhibits extreme variability from near-zero velocities in sheltered pockets to values exceeding 23~m/s over exposed summits. Patch-solver reproduces the majority of these features with errors typically below 8.00~m/s even in the most challenging flow regimes, while Transolver shows visible smoothing of sharp velocity gradients and underprediction of peak accelerations, resulting in more extensive high-magnitude error regions. In particular, at the inflow location $y=2500$, terrain blocking produces large velocity gradients, and Transolver shows substantial bias because it does not explicitly account for the coupling effect of complex terrain. Besides, with increasing altitude, the overall prediction error magnitude of Patch-solver decreases due to the local-attention module compared with Transolver, while the spatial pattern remains broadly similar.

Synthesizing across both representative visualization cases, these results demonstrate that prediction difficulty decreases monotonically with increasing altitude, with near-surface relative L2 errors in complex terrain (Case 2) reaching 0.22-0.30 but declining to 0.03-0.06 by 200~m elevation. Crucially, the performance advantage of dual-attention architectures remains consistently pronounced throughout the near-surface boundary layer (10-150~m), which coincides precisely with the rotor-swept zone of modern utility-scale wind turbines. By 300~m altitude in both cases, where terrain influence becomes negligible and flow patterns homogenize, Transolver and Patch-solver predictions converge to nearly identical accuracy with errors below 1 m/s. This behavior validates that the dual-attention mechanism is not simply adding model capacity indiscriminately, but rather allocating representational power intelligently to the height regime where terrain-flow interactions dominate. The inference speed of 0.73 seconds per forward pass for approximately 300,000 mesh points further compounds these accuracy gains with computational efficiency that transforms the operational paradigm: whereas traditional CFD workflows require hours of simulation time per terrain-inlet configuration, our framework enables the interactive exploration of hundreds of turbine placement scenarios, fundamentally altering the economics of wind resource assessment in complex mountainous terrain.

\subsection{Zero-shot performance for unseen mountains}

\begin{figure}[htbp!]
\centering
\includegraphics[width=\textwidth]{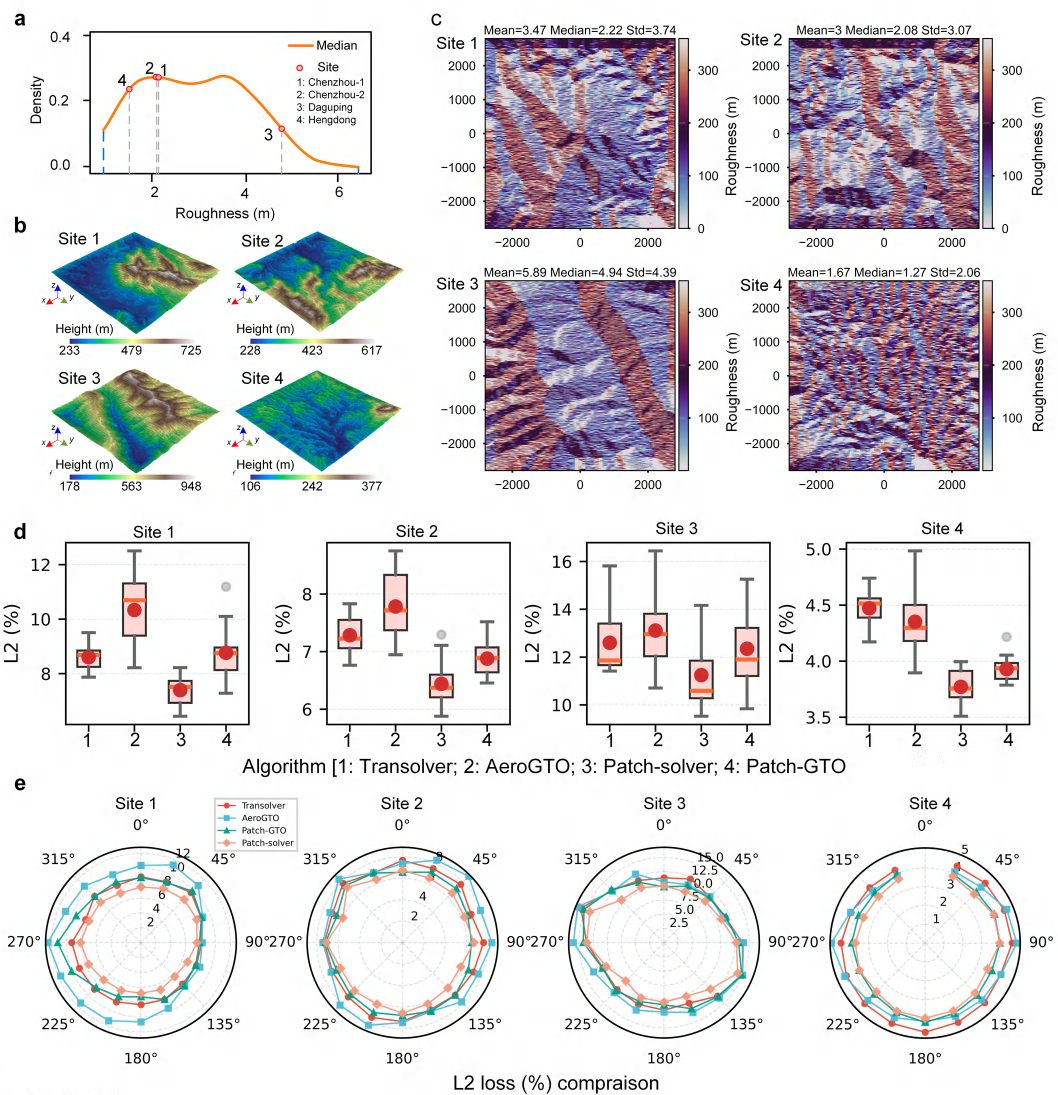}
\caption{\textbf{Zero‑shot inference on four unseen mountainous terrains.} \textbf{a} Kernel–density distribution of surface roughness in the training dataset; vertical dashed markers indicate the per-site median roughness for the four selected locations (Chatou-1, Chatou-2, Daguping, Hengdong). \textbf{b} Digital elevation models for the four test sites. \textbf{c} Spatial roughness maps for four selected mountainous terrains. \textbf{d} Box‑and‑whisker summaries of the relative L2 error (\%) of $U_\text{mag}$ aggregated over all heights and inflow directions for four algorithms (1: Transolver; 2: AeroGTO; 3: Patch‑solver; 4: Patch‑GTO); red dots denote means, boxes show interquartile ranges, and whiskers indicate the non-outlier span. \textbf{e} Direction-resolved relative L2 error (\%) of $U_{\text{mag}}$ across inflow angles (0$^\circ$–360$^\circ$ in 22.5$^\circ$ increments) for each site;  smaller radial extent indicates lower error and reduced directional bias. Note that for site~4 the 0$^\circ$ inflow case was not recorded during data collection.}\label{fig:infer}
\end{figure}

To assess the generalization performance across wind direction and mountain terrain complexity, we conduct a strict zero‑shot evaluation on four mountainous sites (Chatou‑1, Chatou‑2, Daguping, and Hengdong) that are geographically disjoint from the training domain and span markedly different morphologies. Fig. \ref{fig:infer}a positions the four test sites within the training distribution by overlaying their median roughness values onto the kernel density of the training dataset. Sites 1 and 2 (Chatou-1 and Chatou-2) exhibit median roughness values of approximately 2.22 m and 2.08 m, respectively, placing them squarely within the high-density region of the training distribution. In contrast, Site 3 (Daguping) displays a median roughness of 4.94 m, positioning it near the tail of the training distribution, while Site 4 (Hengdong) shows a median roughness of 1.67 m, falling within the lower range but still well-represented in the training data. This distribution analysis establishes that, while all four sites remain nominally within the support of the training distribution, they span a substantial range of morphological complexity that rigorously tests model forecasting generalization capabilities.

The digital elevation models in Fig. \ref{fig:infer}b reveal the distinct topographic characteristics of each site. Sites 1 and 2 present gently undulating terrain with maximum elevation variations ranging from 236 m to 725 m and 226 m to 617 m, respectively, characterized by smooth, continuous slope transitions. Site 3 exhibits dramatically more complex morphology with elevation spanning 178 m to 948 m, featuring multiple prominent peaks, deep intervening valleys, and abrupt ridgeline transitions that create strong three-dimensional flow deflection. Site 4 represents a valley-floor configuration with elevations from 103 m to 377 m, where the terrain forms a natural channel. The corresponding roughness distributions in Fig. \ref{fig:infer}c further illuminate these morphological differences.

% 图改一下 0不要了

Quantitative performance metrics aggregated across all inlet directions and heights are presented in Fig. \ref{fig:infer}d and Table \ref{tab:zero-shot-all}. Across all four sites, Patch-solver and Patch-GTO consistently demonstrate superior generalization capabilities compared to their baseline counterparts. For $U_{mag}$ at Site 1, Patch-solver achieves an average L2 error of 7.40\%, representing a 14.0\% improvement over Transolver's 8.62\% (as detailed in Table \ref{tab:zero-shot-all}), with notably reduced error dispersion as evidenced by the narrower interquartile range. Site 2 exhibits similar trends, with Patch-solver attaining 6.44\% median L2 error versus Transolver's 7.30\%, corresponding to an 11.8\% error reduction. The performance advantage is also pronounced at Site 3, the most topographically complex case: Patch-solver achieves 11.39\% L2 error compared to Transolver's 12.74\%, representing a 10.6\% improvement despite the challenging multi-peak terrain that substantially elevates absolute error levels for all methods. Notably, the box-and-whisker distributions reveal that the Patch-solver not only reduces median error but also compresses the error distribution, indicating more consistent performance across diverse flow scenarios. Site 4, characterized by its valley-floor morphology and lowest absolute roughness, yields the smallest errors across all models, with Patch-solver achieving a 3.78\% L2 error, which represents a 15.9\% improvement over Transolver's 4.49\%, while maintaining the tightest error distribution among all sites. Similarly, Patch-GTO exhibits comparable gains over its baseline (AeroGTO), reinforcing the effectiveness and portability of the proposed solver design.

The directional sensitivity of model performance on $U_\text{mag}$ is systematically plotted in Fig.~\ref{fig:infer}e through polar plots of relative L2 error versus inlet angle for each site. These radar charts visualize how prediction accuracy varies as the wind direction rotates through $360^\circ$ in $22.5^\circ$ increments, with radial distance representing error magnitude. For Site 1, Patch-solver maintains the most compact profile, with its radial extent consistently smaller than that of Transolver, particularly at $90^\circ$ and $270^\circ$, where inlet flow encounters the primary ridgeline. Site 2 shows even less directional sensitivity, with Patch-solver errors of 6\%-7\%, compared with 7\%-10\% for the baselines. Site 3 presents the most pronounced directional dependence, with errors varying from approximately 7\% to 15\% depending on the inlet angle for the proposed Patch-solver. Critically, Patch-solver's advantage persists across all directions at this complex site, maintaining an 8-10\% lower error than Transolver, even at the most challenging inlet angles (approximately $112.5^\circ$ and $292.5^\circ$) where flow obliquely strikes the steepest terrain features. Site 4 exhibits the least directional variation and the lowest absolute errors (3-5\% range), consistent with its valley-channel morphology that produces more predictable flow patterns. Across all sites, the polar plots demonstrate that the dual-attention mechanism's performance advantage is not an artifact of specific inlet configurations but rather a systematic improvement that generalizes across arbitrary flow directions.

These zero-shot results establish that the proposed framework achieves genuine generalization rather than memorization of training-set terrain features. The models successfully transfer to geographically disjoint sites spanning diverse morphological characteristics, from gentle rolling hills to multi-peak massifs to valley channels, while maintaining prediction accuracy within 3-12\% relative L2 error for wind speed magnitude across all evaluated scenarios. The dual-attention mechanism's ability to decompose the prediction task into terrain-local and domain-global components proves particularly beneficial for generalization: local sectional attention adapts to site-specific geometric features without requiring retraining, while global slice attention maintains physically consistent large-scale pressure-velocity coupling. This architectural design enables rapid assessment of wind resources at previously unobserved mountainous locations with accuracy sufficient for preliminary site screening and turbine layout optimization, substantially reducing the computational cost of wind farm development in complex terrain.

\subsection{Wind field prediction with sparse observations}

In practical wind-resource campaigns, reliable observations are usually available at a limited number of locations, for instance from meteorological masts or scanning lidars deployed at several elevations \cite{gao2023optimal,gao2024urban,santos2023development}. We therefore assess how the proposed framework behaves when a small fraction of pointwise wind measurements is provided at the training and inference processes as anchors, and how such sparse priors interact with the dual-attention mechanism. We extend the original wind field prediction problem (Eq.~\ref{eq1: problem}) to incorporate sparse velocity measurements. Given a terrain representation $\mathbf{T}\in \mathbb{R}^{N_\text{s} \times C_\text{in}}$ and query point coordinates $\mathbf{P}\in \mathbb{R}^{N \times 3}$, suppose we have access to sparse velocity observations $\mathbf{V}_{\text{obs}} = \{(\mathbf{p}_j, \mathbf{ws}_j)\}_{j=1}^{M}$ at $M$ monitoring locations, where $M<<N$ and $\mathbf{ws}_j = (u_j, v_j, w_j)$ represent the measured wind velocity at position $\mathbf{p}_j$. The sparse-augmented prediction problem seeks to learn a mapping:
\begin{equation}
\begin{aligned}
\hat{\mathbf{W}} \in \mathbb{R}^{N \times C_\text{in}} = \Phi_\text{sparse}([\mathbf{T}, \mathbf{P}, \mathbf{V}_{\text{obs}}])
\label{eq: sparse_problem}
\end{aligned}
\end{equation}
where $\Phi_{\text{sparse}}$ denotes the sparse-conditioned operator that reconstructs the complete velocity field $\hat{\mathbf{W}} = \{\hat{\mathbf{ws}}_i\}_{i=1}^{N}$ at all $N$ query points by fusing terrain geometry, spatial coordinates, and the sparse observational constraints $\mathbf{V}_{\text{obs}}$. For computational implementation, unmeasured query points are assigned a prior velocity field via nearest-neighbor interpolation:
\begin{equation}
\label{eq:nn_interpolation}
\begin{aligned}
\mathbf {ws}_i^{\text{prior}}
&= \mathbf {ws}_{j^\star}, \\
j^\star
&= \operatorname*{arg\,min}_{\,j\in\{1,\dots,M\}}
\;\bigl\| \mathbf p_i - \mathbf p_j \bigr\|_{2},
\end{aligned}
\end{equation}
where $\mathbf{ws}_j$ represents the measured wind velocity at position $\mathbf{p}_j$. This simple interpolation strategy creates a piecewise-constant velocity field that serves as auxiliary input, effectively treating sparse measurements as input conditions.

As an alternative to nearest-neighbor interpolation, we also consider a Gaussian-process (GP) interpolation scheme to reconstruct the sparse velocity prior from sparse velocity measurements. Specifically, given monitoring locations $\{\mathbf{p}_j\}_{j=1}^{M}$ and their observed velocities $\{\mathbf{ws}_j\}_{j=1}^{M}$, we fit an independent GP regressor for each velocity component ($u$, $v$, and $w$) in the normalized 3D coordinate space. The posterior mean of each fitted GP is then evaluated at the coordinates of all unobserved query points, yielding a smooth prior velocity field over the unknown regions.

In this study, the sparse observations are synthetically generated from the reference CFD solution rather than collected from a field campaign. Specifically, the velocity vectors at selected monitoring locations are directly extracted from the ground-truth CFD results and then perturbed with random noise to emulate sensor measurements. This treatment is motivated by the fact that steady RANS predicts a time-averaged wind field, while practical measurements used for comparison or assimilation are also typically interpreted as time-averaged quantities over a given period. Depending on the overall analysis duration, such averages may correspond to 30-min means or longer-period means. Existing studies have shown that RANS predictions can achieve reasonable agreement with field measurements in terms of these averaged wind characteristics \cite{castorrini2021increasing,cheng2024wind}.

\begin{figure}[htbp!]
\centering
\includegraphics[width=\textwidth]{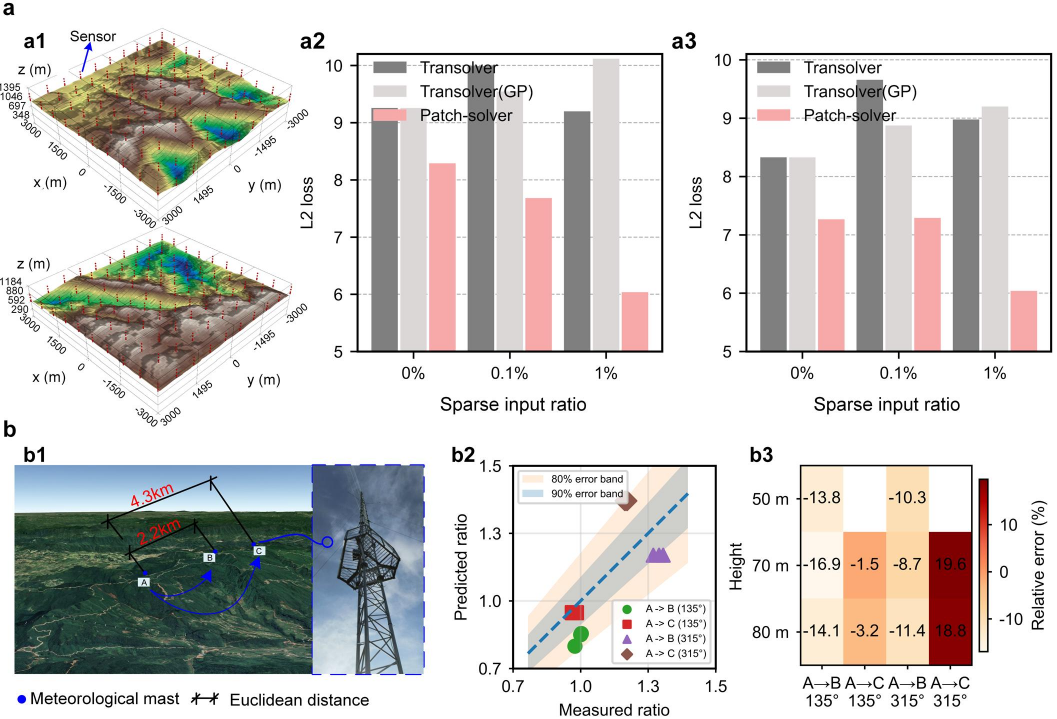}
\caption{\textbf{Two types of application scenarios.} \textbf{a} Sparse-input results on test and zero-shot datasets. \textbf{a1} The spatial distribution of sparse sensors above the terrain with a sparse input ratio of 0.1\% in two typical cases. \textbf{a2} Relative L2 error (\%) of wind speed magnitude $U_{\text{mag}}$ with various sparse input ratios ($M/N=$ 0\%, 0.1\%, 1\%) for Transolver (dark gray), Transolver with Gaussian-process interpolation (light gray) and Patch-solver (pink) evaluated on the test dataset. Sparse measurements are sampled at four standard anemometer heights (50m, 100m, 200m, 300m); unmeasured points inherit velocities from nearest neighbors. \textbf{a3} Same analysis results repeated on the zero-shot dataset. \textbf{b} Real-world inter-mast inference. \textbf{b1} Map of the three meteorological masts (A, B, C) and Euclidean distances between them in a practical terrain. \textbf{b2-b3} Predicted and measured wind-speed ratios at multiple heights (above the terrain) and two representative wind directions ($135\degree$ and $315\degree$).}\label{fig:sparse}
\end{figure}

To mimic realistic sparse-measurement scenarios, we extract ground-truth velocity vectors at monitoring points positioned at four standard anemometer altitudes: 50~m, 100~m, 200~m, and 300~m above the local terrain surface, as depicted in Fig.~\ref{fig:sparse}a1. At each altitude plane, we sample monitoring locations following a spatially uniform distribution, ensuring unbiased coverage across the terrain. We evaluate two sparse input ratios: $M/N$=0.1\% (approximately 300 monitoring points) and 1\% (approximately 3000 monitoring points). The sparse-observation pipeline is kept identical in both training and test stages: monitoring points are sampled at the same predefined height levels and spatial densities. Both Transolver and Patch-solver are evaluated on two datasets: the test set, consisting of terrain configurations drawn from the same distribution as the training data, and the zero-shot dataset, comprising four geographically distinct unseen mountainous sites spanning diverse inflow angles.

Fig.~\ref{fig:sparse} presents the relative L2 error for wind speed magnitude with various sparse input ratios for both models across the two evaluation datasets. In the baseline scenario without observational data on the test set (Fig.~\ref{fig:sparse}a2), Patch-solver achieves an 8.3\% L2 error compared to Transolver's 9.2\%, maintaining the approximately 10\% performance advantage established in previous sections. When incorporating sparse measurements at just 0.1\%, Transolver achieves a 10.0\% error, a counterintuitive 8\% increase from baseline (without sparse input), suggesting that its global attention mechanism struggles to effectively localize and leverage such sparse constraints. In contrast, Patch-solver improves to 7.7\%, representing a 7.2\% relative error reduction from its baseline and demonstrating superior capacity for sparse data assimilation. The performance gap widens substantially as measurement density increases to 1\%: Patch-solver achieves a 6.0\% error while Transolver reaches 9.2\%—essentially unchanged from baseline. Remarkably, with merely 1\% observational coverage, Patch-solver achieves a 27.7\% error reduction relative to its pure-geometry baseline (from 8.3\% to 6.0\%) and maintains a 34.8\% accuracy advantage over Transolver. This dramatic improvement indicates that even extremely sparse measurements, when properly assimilated through the dual-attention architecture, can substantially enhance prediction fidelity.

% 取点的示意图 a 三维画一下。

On the zero-shot mountainous sites (Fig.~\ref{fig:sparse}{a3}), Patch-solver retains a clear advantage over Transolver at every sparse ratio and, crucially, is the only model that converts additional sparse measurement data into a meaningful accuracy gain. Without sparse measurement input, Transolver records a relative L2 error of 8.33\%, while Patch-solver is lower at 7.27\% (a 12.76\% advantage). Introducing a 0.1\% sparse input ratio degrades Transolver to 9.66\%, which represents a 15.95\% increase over its own baseline. However, the Patch-solver remains essentially unchanged at around 7.2\%. At a 1\% sparse input ratio, Transolver recovers only partially to 8.98\% (still 7.80\% above baseline), but Patch-solver improves sharply to 6.04\%, a 16.89\% reduction relative to its baseline (without sparse data input). This corresponds to Patch-solver’s relative advantages of 12.76\%, 24.51\%, and 32.75\% at 0\%, 0.1\%, and 1\% sparse input ratios compared with Transolver, respectively. These results indicate that, on previously unseen terrain, sparse in-situ measurements tend to destabilize a purely global-attention solver (such as Transolver), whereas the dual-attention design in Patch-solver first localizes anchor information within voxel neighborhoods and then reconciles it globally, turning very small observations into domain-wide improvements without any site-specific re-training.

To disentangle the effect of prior construction from that of network architecture, we further replace the nearest-neighbor interpolation with Gaussian-process (GP) interpolation as an alternative sparse-prior construction strategy. As shown in Fig.~\ref{fig:sparse}a2-a3, GP interpolation does not yield a consistent improvement over nearest-neighbor interpolation. Although it produces a smoother prior field and can be beneficial at lower sparse-input ratios, its performance gain remains limited and may even deteriorate at higher sparse-input ratios. A likely reason is that, except at the truly observed locations, the remaining values are still interpolated pseudo-observations rather than real measurements. Once these imperfect dense priors are propagated through a purely global-attention architecture such as Transolver, the accumulated interpolation errors may interfere with the reconstruction of the true flow field.

% 结果也表明，尽管在低稀疏输入比下效果较好，但是提升稀疏输入比效果也挺一般的。这这主要是因为除了稀疏观测位置的数据外，其他数据都是假数据，全局注意力很容易让这些假的数据影响预测效果。这个收益远不如模型改变带来的收益稳定。由此也证明了模型的有效性。

% This comparison serves as an ablation on prior design. We observe that improving the prior generally benefits sparse-conditioned prediction, confirming that the quality of the assimilated prior field is an important factor. However, the gain obtained by Patch-solver remains larger and more consistent than that of Transolver under the same sparse-observation setting, indicating that the performance improvement cannot be attributed to the prior alone. Instead, the results suggest that Patch-solver is better able to exploit informative sparse priors once they are injected into the model.

The superior sparse-data assimilation capability of Patch-solver stems directly from its dual-attention architectural design, which naturally decomposes the reconstruction task into complementary local and global inference pathways. Within each voxel partition, the local attention mechanism enables query points to selectively attend to nearby measurement locations, effectively treating sparse observations as soft boundary conditions that constrain the solution locally. When a voxel contains one or more measurement points, the local attention weights concentrate on these observed velocities, rapidly adapting predictions within that spatial region to align with empirical data. This locality-preserving mechanism ensures that measurement information is exploited for targeted accuracy improvements rather than being diffused across the entire domain. Simultaneously, the global attention pathway propagates measurement-derived constraints across the entire computational domain through long-range slice tokens, ensuring that local corrections induced by sparse observations remain physically consistent with large-scale pressure-velocity coupling and terrain-induced flow patterns. This global coordination prevents the formation of spurious discontinuities at voxel boundaries and maintains adherence to underlying conservation laws even when incorporating empirical data.

\subsection{Real-world inter-mast wind speed inference}

To further examine the practical applicability of the proposed framework, we consider a real-world mast-to-mast wind inference task based on field measurements from an actual complex mountainous site instrumented with multiple meteorological masts. Such a scenario is common in wind-resource assessment, where only limited on-site observations are available and the wind condition at a target location must be estimated from a nearby reference mast due to practical measurement constraints, such as limited instrumentation or asynchronous observation periods across sites. Here, mast A is treated as the reference mast, while masts B and C are regarded as target locations, as shown in Fig.~\ref{fig:sparse}b1. The annual-mean horizontal wind speed, defined as $U_\mathrm{h}=\sqrt{u^2+v^2}$, at several representative heights (50~m, 70~m, and 80~m) at two target masts (B and C) needs to be inferred from the observation of a reference mast (A). Rather than comparing only absolute wind speeds, we focus on inter-mast wind speed ratios, such as $U_\mathrm{h}^{\mathrm{B}}/U_\mathrm{h}^{\mathrm{A}}$ and $U_\mathrm{h}^{\mathrm{C}}/U_\mathrm{h}^{\mathrm{A}}$, because these ratios directly characterize the terrain-modulated transfer relationship from the reference mast to target masts.

In the inference procedure, only the site terrain geometry with a specific inflow angle is provided to the trained model. The model predicts the full wind field over the site, after which the horizontal wind speeds at the locations of masts A, B, and C are extracted at the corresponding heights. The predicted inter-mast ratios are then compared against the measured annual-mean ratios obtained from the actual mast observations. As illustrated in Fig.~\ref{fig:sparse}b2 and b3, the predicted ratios generally agree well with the measurements across different mast pairs and heights. The average absolute relative error is 11.83\%, with all cases remaining below 20\%, and several combinations achieving errors below 3\%. This indicates that the model is capable of capturing the dominant terrain-induced transfer relationships between masts. Although this experiment is limited to a single real-world site and two representative wind directions, the results nevertheless demonstrate the potential of the proposed framework for practical wind-resource assessment and mast-to-mast wind extrapolation in complex terrain.

\subsection{Interpretability analysis}

We assess how the dual–attention operator integrates information across space and scales by analyzing the Shannon entropy of its learned attention distributions. For each query point $i$ within a segmented patch of size $n$, let $\mathbf A\in\mathbb R^{n\times n}$ denote the head-averaged, patch-internal attention matrix (obtained by averaging per-head softmax weights), where $\mathbf{A}_{i,j}\!\ge\!0$ quantifies the attention from neighbor $j$ to point $i$. To ensure numerical robustness, we normalize each row to a probability vector:
\begin{equation}
\begin{aligned}
\mathbf P_{i,j}
=\frac{\mathbf A_{i,j}}{\sum_{k=1}^{n}\mathbf A_{i,k}},\qquad
\sum_{j=1}^{n}\mathbf P_{i,j}=1
\end{aligned}
\end{equation}
By construction, the row $\mathbf P_{i:}$ gives the exact attention coefficients of neighbors within the current patch used to update point $i$.

The local attention entropy ($H_\text{local}$) for point $i$ is then defined as:
\begin{equation}
\begin{aligned}
H_{\text{local},i} \;=\; - \sum_{j=1}^{n} \mathbf P_{i,j}\,\log\!\big(\mathbf P_{i,j})
\label{eq:hlocal_def}
\end{aligned}
\end{equation}
which describes how spread the attention is within the patch: small values mean the point relies on only a few neighbors (focused), large values mean it uses many neighbors with similar weights (spread out).

In parallel, the global pathway assigns each point a head-averaged soft vector $\mathbf{w}_i\in[0,1]^G$ over $G$ slice tokens with $\sum_{g}w_{i,g}=1$. The global slice entropy can then be represented as:
\begin{equation}
\begin{aligned}
H_{\text{global},i} \;=\; - \sum_{g=1}^{G} w_{i,g}\,\log\!\big(w_{i,g})
\label{eq:hglobal_def}
\end{aligned}
\end{equation}
where a small value represents specialization to a few global modes, whereas larger values indicate reliance on multiple domain–wide patterns.

\begin{figure}[htbp!]
\centering
\includegraphics[width=0.99\textwidth]{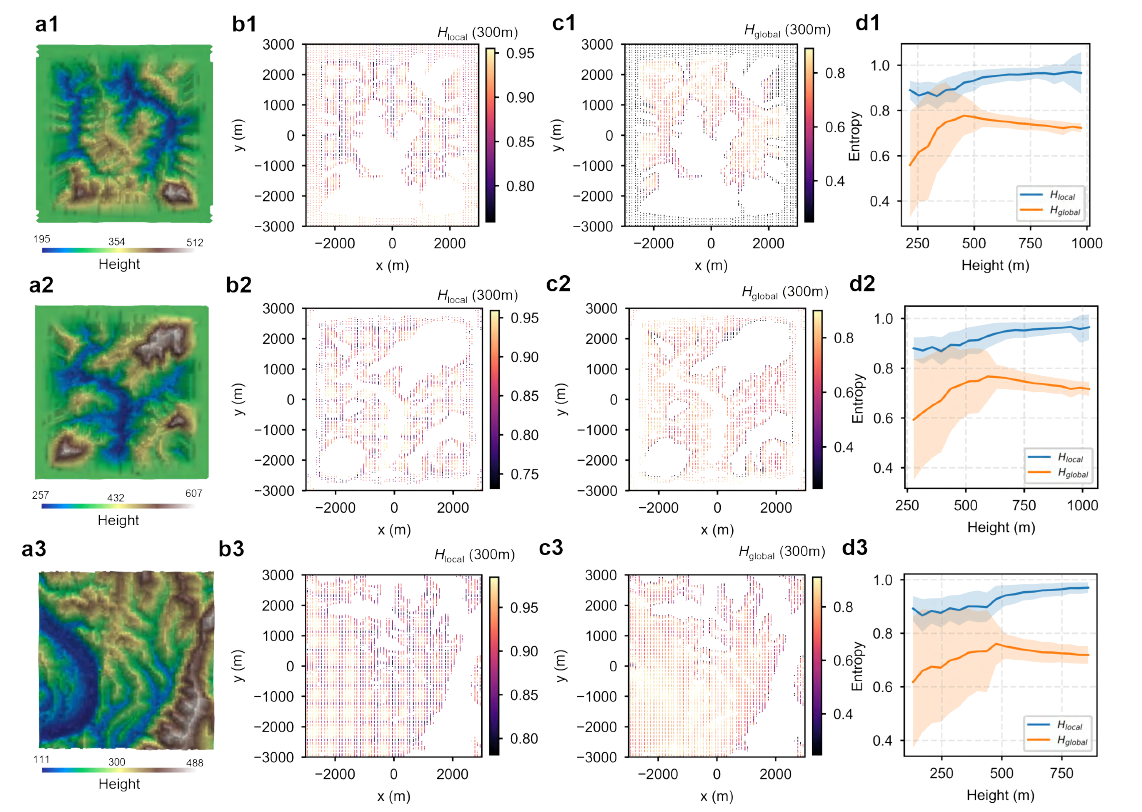}
    \caption{\textbf{Interpretability analysis of dual–attention proposed in this paper.}
    \textbf{a1–a3} Schematic diagram of mountain terrain for three test scenes.
    \textbf{b1–b3} Normalized local entropy $H_{\text{local}}$ at 300\,m height.
    \textbf{c1–c3} Normalized global entropy $H_{\text{global}}$ at 300\,m height. 
    \textbf{d1–d3} The variation of local and global entropies (mean and standard deviation) with height. In subplots \textbf{d1–d3}, solid lines denote the mean values, and the shaded band indicates $\pm 1$ standard deviation around the mean values.}
\label{fig:interpre}
\end{figure}

Throughout this section, global and local entropies are reported in their scale–invariant form, i.e., $H_{\text{local}}/\log(n)$ and $H_{\text{global}}/\log(G)$, so that values lie in $[0,1]$ and are comparable across patches and slice counts. In this normalization, larger values indicate more even mixing and smaller values indicate sharper concentration.Fig.~\ref{fig:interpre} shows the entropy variation of three representative terrains with various complexities.

Fig.~\ref{fig:interpre}a1-a3 present the digital elevation models that define the terrain geometry. The maps of normalized local entropy at the absolute elevation $z = 300$~m ( Fig.~\ref{fig:interpre}b1-b3) reveal that a low entropy value is generated at the position where the terrain height changed, but a relatively high entropy value is maintained at places where the height remained approximately the same as neighbors. Besides, for the normalized global entropy (Fig.~\ref{fig:interpre}c1-c3), it is revealed that, within complex terrain, points tend to combine several slice tokens rather than committing to a single one, yielding higher entropy than in gentler areas where assignments concentrate on a limited subset of slices. The curves in Fig.~\ref{fig:interpre}d1-d3 quantify the vertical trend versus absolute elevation $z$ and show a consistent increase for both quantities: near the ground, the local attention relies on only a few neighbors and the global assignment specializes to a small set of terrain-conditioned slices, which keeps both entropies low, However, with the increase of the height, the flow is more uniform and the learned slices behave more interchangeably, so both the local and the global normalized entropies rise with height. 

The entropy analysis provides crucial insights into the adaptive nature of the proposed dual-attention mechanism and validates several key architectural design choices. First, the spatial heterogeneity of local entropy demonstrates that the model automatically allocates representational capacity based on local terrain complexity: regions with abrupt topographic changes trigger focused, low-entropy attention patterns that concentrate on immediately adjacent neighbors, while smoother areas maintain higher entropy and distribute weights more evenly predominantly at locations with low terrain roughness. This adaptive behavior confirms that the local sectional attention is not simply averaging neighboring points indiscriminately but rather intelligently identifying geometric discontinuities that require specialized treatment—a capability that purely global attention mechanisms lack.

Second, the vertical stratification of both entropy measures reveals a physically interpretable transition from terrain-dominated to free-stream regimes. Near the surface, the combination of low local entropy (sharp neighbor selection) and low global entropy (specialization to terrain-conditioned slice tokens) indicates that the model recognizes the dominant role of boundary-layer physics and local geometric constraints. As elevation increases and terrain influence wanes, both entropy measures rise monotonically, signifying that the model smoothly transitions to a more homogeneous flow representation where long-range correlations become equally important and local geometric features lose their dominance. This vertical gradient in attention behavior directly mirrors the known physics of atmospheric boundary-layer development over complex terrain, providing strong evidence that the learned representations are physically meaningful rather than spurious statistical artifacts.

\section{Discussion}
\label{sec:dis}

This work proposes a transformer-based dual-attention neural operator framework for real-time prediction of three-dimensional wind fields over complex mountainous terrain, addressing fundamental limitations of both mesh-based DL methods and computationally expensive CFD simulations. Through systematic validation across diverse topographic configurations and flow conditions, we demonstrate that this dual-attention design is a general architectural principle by instantiating it on two representative operator backbones.

Across these two instantiations, the proposed dual-attention neural operators achieve approximately 8\% relative L2 error for wind speed magnitude averaged across the entire 3D wind field on the test set, while delivering sub-second inference for around 360,000 query points, representing roughly a 10\% improvement over strong neural operator baselines. The performance gains are most pronounced in the near-surface boundary layer (10–150~m), precisely the rotor-swept zone critical for turbine siting. As elevation increases and terrain influence diminishes, prediction errors decrease monotonically, with all models converging to comparable accuracy ($<$2\% error) at 300~m where free-stream conditions prevail. This vertical stratification demonstrates that the dual-attention architecture intelligently allocates representational capacity to height regimes where terrain-flow coupling is strongest. 

The framework exhibits robust generalization capabilities essential for operational deployment. Zero-shot evaluation on four geographically disjoint mountainous sites spanning diverse morphologies yields 4–12\% relative L2 errors without any site-specific retraining, with consistent performance advantages maintained across all 16 inflow directions tested. Furthermore, the framework's capacity to assimilate sparse in-situ measurements (0.1–1\% spatial coverage) proves particularly valuable: while baseline models struggle to leverage such sparse constraints effectively, Patch-solver converts minimal observational data into substantial accuracy gains, achieving 32.75\% lower error than the strong baseline model with just 1\% measurement coverage on unseen terrains. In addition, the framework also demonstrates practical utility in a real-world mast-to-mast wind inference task. Using terrain geometry alone as input, the predicted inter-mast wind speed ratios agree with annual-mean measurements at an average absolute relative error of 11.83\%. Finally, entropy-based analysis provides an interpretable lens: the learned attention becomes increasingly diffuse with height, mirroring the physical transition from terrain-dominated boundary-layer flow to more homogeneous free-stream regimes, which supports that the model captures meaningful structure rather than relying on spurious shortcuts.

Despite these advances, several limitations warrant consideration for future work. First, the current framework is trained exclusively on CFD-simulated data, and bridging the gap to real-world observations remains an open challenge. Since CFD simulations rely on simplified boundary conditions and turbulence models, discrepancies may exist between simulated and actual atmospheric flows. In practical applications, incorporating in-situ monitoring data through a correction model may help reduce this gap and further improve predictive reliability. Second, incorporating physics-informed loss terms into the operator learning framework may further constrain the prediction of these secondary velocity components and improve overall accuracy. In addition, when sufficient computational resources are available, training separate models or dedicated prediction branches for each velocity component may further improve the accuracy of the secondary components by allowing the network to learn component-specific flow features more effectively.

Moreover, the present study focuses on steady-state wind fields under neutral atmospheric stability, whereas practical wind resource assessment increasingly demands time-varying predictions that capture diurnal cycles, transient weather events, and turbulent fluctuations relevant to fatigue loading and grid integration. Extending the framework to unsteady flows, potentially through temporal attention mechanisms or coupling with mesoscale model outputs, represents a natural and important direction.

\section{Method}
\label{sec:method}
We instantiate the proposed dual-attention principle in two complementary neural operator architectures that represent distinct paradigms in geometric deep learning: Patch-solver, a Transformer-based implementation that operates directly on point coordinates through global slice attention and local voxel-based sectional attention; and Patch-GTO, a graph-operator variant that explicitly constructs edge connectivity by combining the original mesh edges with additional edges established via KNN search, and processes messages through attention-based aggregation modules. Both architectures share the core dual-attention design—decomposing the solution operator into local terrain-confined and global domain-wide pathways—but differ in how they encode spatial relationships: Patch-solver relies on implicit geometric features (spatial coordinates, SDF field) processed through attention, while Patch-GTO leverages explicit graph topology to aggregate information from spatially adjacent neighbors. In the following subsections, we detail the Patch-solver architecture and its key components, with Patch-GTO adopting an analogous dual-attention integration within the graph-operator framework.

\subsection{Patch-solver}
The neural operator framework processes unstructured point clouds sampled from the atmospheric domain above mountainous terrain, where each query point requires geometric context to infer its local velocity field. For each query point $\mathbf{p}_i = (x_i, y_i, z_i)$ in the computational domain, we construct a 7-dimensional feature vector input $X_i=[x_i, y_i, z_i, \hat{n}_x, \hat{n}_y, \hat{n}_z, d_i]$ that encodes both its absolute spatial location and its geometric relationship to the underlying terrain surface. The SDF contributes four geometric descriptors: the scalar distance value $d_i$ representing the minimum Euclidean distance to the nearest terrain surface, and three components of the unit surface normal $\hat{\mathbf{n}}_i = (\hat{n}_x, \hat{n}_y, \hat{n}_z)$. A detailed description of the input configuration and training details can be found in Appendix \ref{traing config}.

\begin{figure}[htbp!]
\centering
\includegraphics[width=\textwidth]{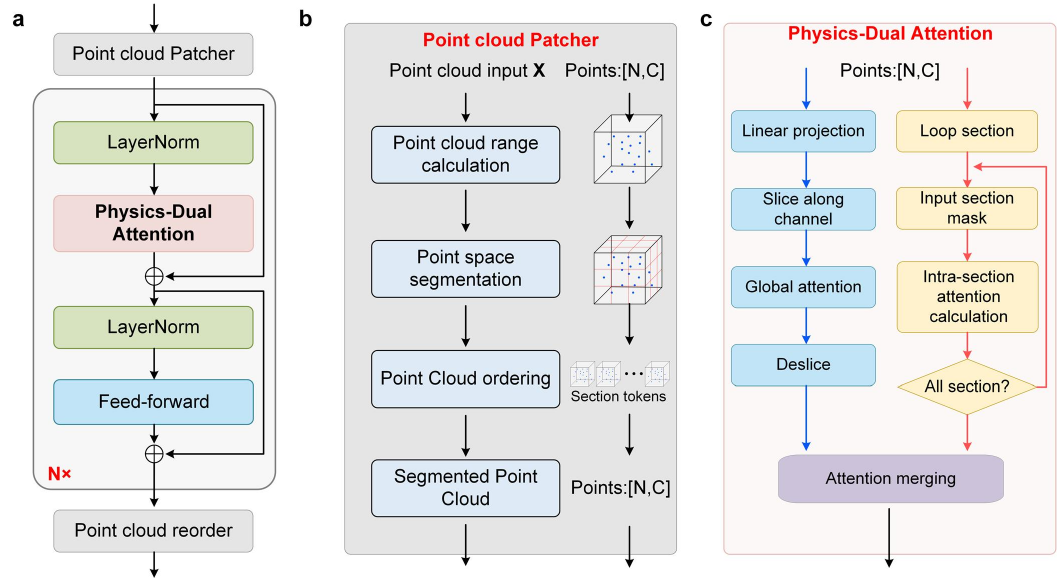}
\caption{\textbf{Architecture of proposed Patch-solver.} \textbf{a} The basic architecture of the proposed Patch-solver model. \textbf{b} Description of the Point Cloud Patcher module. \textbf{c} Description of the physics-Dual Attention. }\label{fig:1}
\end{figure}

The built feature input ($\mathbf{X}$) is handled by the Point Cloud Patcher module (PCPM), Physics Encoder module (PEM), and Point Cloud Reorder (PCR) technology in sequence, as depicted in  Fig. \ref{fig:1}. PCPM is used to patch the irregular CFD point clouds and save the input order. Then, the PEM is used to map the ordered point clouds to the global wind velocity field of the complex mountain terrain, where the primary solver mechanism, Physics–Dual Attention module (PDAM), is used to capture the global and local interactions inside the input point clouds. The main architecture of the PEM is inspired by the Transformer encoder architecture and the Transolver methodology to compute global attention with high efficiency.

% 为了便于后续的attention计算，我们在

\subsection{Point Cloud Patcher module}

The PCPM organizes the input points into local patches for efficient attention computation. Importantly, this operation does not resample the irregular point cloud onto a regular grid. Instead, it only uses a regular voxel lattice as an indexing structure to assign patch identifiers, while preserving the original point coordinates and associated features throughout the prediction process. Specifically, given an irregular CFD point cloud $\mathbf{P} \in \mathbb{R}^{N\times C_{\mathrm{in}}}$ with point coordinates $\mathbf{X}_i$, PCPM partitions the bounding box of the point cloud into a regular lattice of $(n_x,n_y,n_z)$ voxels and assigns each point a unique patch identifier:
\begin{equation}
\operatorname{id}_i
=
\lfloor (\mathbf X_i-\mathbf X_{\min})\odot(n_x,n_y,n_z)/(\mathbf X_{\max}-\mathbf X_{\min}+10^{-6}) \bigr\rfloor
\label{eq1: 1}
\end{equation}
where symbol $\odot$ indicates element-wise multiplication. The function $\lfloor\cdot\rfloor$ applies the floor operation to each component of the vector, returning the largest integer not greater than the argument. The constant $10^{-6}$ is a small numerical stabilizer that prevents division by zero when $\mathbf X_{\max}-\mathbf X_{\min}$ happens to contain a zero in any coordinate direction. $(n_x,n_y,n_z)$ denotes the voxel resolution along the $x$-, $y$- and $z$-axes, respectively.

Then, PCPM  performs a stable sort on $\{\operatorname{id}_i\}$ to produce the reordered tensor $\mathbf X_{\text{sorted}}\in\mathbb R^{N\times C_{\mathrm{in}}}$.  
The inverse permutation $\pi^{-1}$ is stored so that the final prediction can be mapped back to the original ordering without numerical drift.

\subsection{Physics Encoder module}
Within PEM, the reordered sequence is first linearly projected to hidden representation $\mathbf{Z} \in \mathbb{R}^{N\times H}$, followed by a layer normalization to ensure numerical stability and scale invariance. This normalized feature is then fed into the Physics–Dual Attention Modules (PDAMs), which propagate information both within individual voxels and across the entire spatial domain. the PDAM simultaneously captures small-scale turbulent eddies tightly confined to the terrain surface and the large-scale pressure–velocity coupling that extends across the entire mountainous domain. To preserve representational fidelity and enhance gradient flow in deep architectures, a residual connection is introduced by directly adding the output of PDAM to the input $\mathbf{Z}$. This residual-enhanced representation is further refined via another layer normalization and a position-wise feed-forward multilayer perceptron (MLP). The PEM is structured as a multi-layer deep neural architecture, where the output of one PDAM block serves as the input to the next. The entire forward propagation can be formally expressed as:
\begin{equation}
\mathbf {MZ}^{\ell+1}=\mathbf Z^{\ell}+\text{PDAM}(\text{LayerNorm}(\mathbf Z^{\ell}))
% \sigma(g)\operatorname{LN}\!\bigl(\mathbf Y^{\mathrm{local}}\bigr)+\bigl(1-\sigma(g)\bigr)\operatorname{LN}\!\bigl(\tilde{\mathbf F}\bigr),
\label{pem1}
\end{equation}
\begin{equation}
\mathbf Z^{\ell+1}=\text{Feedforward}(\text{LayerNorm}(\mathbf {MZ}^{\ell+1}))+\mathbf {MZ}^{\ell+1}
\label{pem2}
\end{equation}
where $ Z^{\ell+1}$ denotes the hidden feature matrix that reaches the $\ell$-th PEM layer. LayerNorm($\cdot$) rescales and recentres each feature channel to zero mean and unit variance. Feedforward() is the multilayer perceptron consisting of two linear projections separated by a GELU activation and dropout.

\subsection{Physcis-Dual attention}

The dual-attention operator is the key to balancing efficiency and physical fidelity for the proposed PDE solver. The input representation are sent to two complementary attention kernels (global slice attention and local sectional attention) in parallel and fusing their outputs through a learnable gate. 
% The global slice attention, inspired by Transolver model, operates on a greatly compressed set of tokens and remembers global interactions that are essential for capturing global wind speed filed distribution. 

\paragraph{(i) Slice-based global physics attention}
Following the physics attention kernel of Transolver, each head softly compresses the whole point set into $G (\ll N)$ slice tokens.For one attention head $h$, the slice weight of token $g$ for mesh point $i$ is:
\begin{equation}
w_{i,g}^{h} =
\operatorname{Softmax}\!\Bigl(
    (\mathbf{F}_{i}\mathbf{W}^{h}_{\text{slice}})_g / \tau_{h}
\Bigr)\qquad
\label{gla1}
\end{equation}
\begin{equation}
\sum_{g=1}^{G} w_{i,g}^{h}=1
\label{gla2}
\end{equation}
where $\mathbf{F} \in \mathbb{R}^{N \times H_\text{d}}$ is calculated by projecting the hidden size ($H$) of input representation to the integer multiple of the number of attention heads ($H_\text{n}$). $\tau_{h}$ is a learnable temperature and $\mathbf{W}^{h}_{\text{slice}}\in\mathbb{R}^{G \times H_{\!d}}$ is the projection weights that map a point feature to slice affinities for head $h$.

Then, the weights are aggregated in the point space, and the $g$‑th slice token can be calculated as:
\begin{equation}
\mathbf{s}_{g}^{h} =
\frac{\sum_{i=1}^{N} w_{i,g}^{h} \,\mathbf{F}_{i}}
     {\sum_{i=1}^{N} w_{i,g}^{h}+10^{-5}}
     \;\in\mathbb{R}^{1 \times H_{\!d}}
\label{eq:slice_token}
\end{equation}

Thus, self‑attention can be computed in the compressed $G$‑token space without large computation afford:
\begin{equation}
\tilde{\mathbf{s}}_{g}^{h}
=
\sum_{g'=1}^{G}
\operatorname{Softmax}\!\Bigl(
\tfrac{ (\mathbf{s}_{g}^{h}\mathbf{W}_{Q}^{h})
        (\mathbf{s}_{g'}^{h}\mathbf{W}_{K}^{h})^{\!\top}}
      {\sqrt{H_{\!d}}}\Bigr)
\,(\mathbf{s}_{g'}^{h}\mathbf{W}_{V}^{h})
\label{eq:attn}
\end{equation}

Finally, the refined global context is desliced back to every point and inverse projected to the original hidden size ($H$),
\begin{equation}
\mathbf{G}_{i}^{h}
=
\sum_{g=1}^{G} w_{i,g}^{h}\,\tilde{\mathbf{s}}_{g}^{h}
\label{eq:deslice}
\end{equation}

After head concatenation, we obtain the domain-wide correlations:
\begin{equation}
\mathbf G=\operatorname{concat}_{h=1}^{H_\text{n}}\mathbf G^{(h)}
\in\mathbb R^{N\times H}
\label{eq:ALLHEAD}
\end{equation}

\paragraph{(ii) Local sectional attention} The PCPM has already grouped the $N$ points into $M=n_xn_yn_z$ contiguous voxel segments
$\{\,\mathcal{S}_k\,|\,k=1,\dots,M\}$ with cardinalities $|\mathcal{S}_k|=N_k$ and $\sum_{k}N_k=N$. For every patch $s_k$ and every head $h\in\{1,\dots,H\}$, query, key and value projections are formed as:
\begin{equation}
\mathbf Q_k^{(h)}=\mathbf F_{s_k}\mathbf W_Q^{(h)},\quad
\mathbf K_k^{(h)}=\mathbf F_{s_k}\mathbf W_K^{(h)},\quad
\mathbf V_k^{(h)}=\mathbf F_{s_k}\mathbf W_V^{(h)}
\in\mathbb R^{N_k\times H_\text{d}}
\label{eq:ALLHEAD-local}
\end{equation}

Here, scaled dot-product attention is evaluated \emph{exclusively} inside the voxel:
\begin{equation}
\mathbf A_k^{(h)}=\operatorname{softmax}
\!\Bigl(
\mathbf Q_k^{(h)}\mathbf K_k^{(h)\!\top}/\sqrt{d_h}
\Bigr),\qquad
\mathbf L_{s_k}^{(h)}=\mathbf A_k^{(h)}\mathbf V_k^{(h)}
\label{eq:ALLHEAD-local}
\end{equation}

Concatenating all heads and all sections yields the local response:
$\mathbf L=\operatorname{concat}_{h,k}\mathbf L_{s_k}^{(h)}
\in\mathbb R^{N\times H}$,
which preserves the sharp, anisotropic flow features that are strictly
confined to each voxel neighborhood.

\paragraph{(iii) Gated fusion and residual update}
A single learnable scalar $g\!\in\!\mathbb R$
balances the two streams:
\begin{equation}
\mathbf Y
=\sigma(g)\,\operatorname{LN}(\mathbf L)
+\bigl(1-\sigma(g)\bigr)\,\operatorname{LN}(\mathbf G)
\label{eq:FINAL}
\end{equation}
where $\sigma(\cdot)$ is the Sigmoid function
and LN denotes layer normalization applied channel-wise.
The PDAM output is finally added to the input to form a
residual branch:
\begin{equation}
\mathbf Z^{\prime}=\mathbf Z+\mathbf Y
\label{eq:FINAL}
\end{equation}

%输入的点云首先会进入我们的point cloud patcher 模块，考虑在空间中的位置关系，我们将其按照区域分割为 M1*M2*M3个子块，并计算为唯一的位置掩码。随后把点云的点所属的位置掩码进行计算。我们按照从小到大将位置掩码进行排列，并把空间中的点按照位置掩码进行重排序。注意，这时候，每个点原本的坐标索引我们也保留了下来，方便后续进行逆转回原来点的位置，以保证损失的正常计算。

\subsection{Loss function}

We minimize a two-term loss that combines a terrain boundary penalty with a data-fidelity term, as described in Eq. \ref{loss}. The first term enforces a zero-velocity condition on the terrain by penalizing the velocity magnitude at surface points $i \in \Omega_\text{te}$, and the second term is the Mean-Squared Error (MSE) between targets.
\begin{equation}
\begin{aligned}
% \mathcal{L} = \mathcal{L}_{\text{terrain}} + \mathcal{L}_{\text{MSE}}
\mathcal{L} = \mathcal{L}_{\text{terrain}} + \mathcal{L}_{\text{MSE}}=\sum_{i \in \Omega_{\text{te}}} \|\mathbf{\hat{y}}_i\|^2 + \frac{1}{N}\sum_{j=1}^{N} \|\mathbf{y}_j - \mathbf{\hat{y}}_j\|^2
\label{loss}
\end{aligned}
\end{equation}
where $\mathbf{\hat{y}}_i$ represents the predicted wind velocity on the terrain surface. $\mathbf{\hat{y}}_j$ and $\mathbf{{y}}_j$ are the predicted and actual wind velocity vectors, respectively.

\section*{Data availability}
Due to confidentiality agreements with the industrial partner, the full dataset generated and analyzed during this study is not publicly available.  However, a representative subset of the data has been deposited on GitHub to enable reproducibility of the main findings.

\section*{Code availability}
The source code used to train and run the neural operator models in this study
is available on GitHub.

\section*{Acknowledgements}
S.Z. Cai acknowledges support from the Zhejiang Provincial Natural Science Foundation of China (Grant No. LZ24F030003) and the Fundamental Research Funds for the Central Universities.

\section*{Author contribution statements}
\textbf{Yujia Zhang}: Conceptualization, Investigation, Data curation, Formal analysis, Visualization, Software, Writing - review\&editing, Methodology;
\textbf{Jiaxi Qi}: Investigation, Data curation, Formal analysis, Visualization, Software;
\textbf{Ruiyan Chen}: Conceptualization, Resources, Data curation;
\textbf{Yong Liu}: Conceptualization, Resources, Data curation;
\textbf{Yuzhou Zhang}: Conceptualization, Data curation;
\textbf{Lyulin Kuang}: Conceptualization, Data curation;
\textbf{Rita Zhang}: Conceptualization, Funding acquisition;
\textbf{Shengze Cai}: Conceptualization, Investigation, Funding acquisition, Writing - review\&editing, Project administration.

\section*{Competing interests}
The authors declare no competing interests.

% \section*{Appendix}

 \bibliographystyle{elsarticle-num} 
 \bibliography{cas-refs.bib}

\clearpage
 
\begin{appendices}

\counterwithin{figure}{section}   % 让图编号跟随附录章节（如 A.1, A.2）
\counterwithin{table}{section}    % 让表编号跟随附录章节（如 A.1, A.2）
\counterwithin{equation}{section} % 如果需要公式也重编号

\section{Numerical simulation of flow over complex terrain}
\label{OPENfoam}

This study uses a modified $k$-$l$ turbulence model \cite{chen2024kl} specifically adapted for large-scale flows over complex terrain. The model has been validated against field measurements from the Bolund experiment~\cite{berg2011bolund,bechmann2011bolund} and from an operational wind farm in southern China. The core analysis framework is based on steady Reynolds-averaged Navier--Stokes (RANS) simulations, through which the wind field distributions over complex terrain are computed under 16 different inflow directions. 

It is worth noting that, as a steady-state modeling approach, RANS analysis typically corresponds to the mean wind-field distribution over a certain averaging period in real measurements, such as several hours, days, or even longer \cite{castorrini2021increasing,cheng2024wind}, depending on the practical application scenario. On the other hand, RANS simulations can provide physically consistent wind-speed relationships between arbitrary locations across the terrain, which is particularly valuable for subsequent engineering applications, such as inferring the actual wind speed at any potential locations from measurements collected by meteorological masts. The detailed numerical setup is described below.

\subsection{Governing equations and solver}

The wind field over complex terrain is computed by solving the three-dimensional, steady-state, incompressible RANS equations with the SIMPLE algorithm. The governing equations are:
\begin{equation}
\begin{aligned}
  \nabla \cdot \mathbf{U} &= 0,
  \label{eq:mass_app}\\
  \nabla \cdot \bigl(\mathbf{U}\otimes\mathbf{U}\bigr)
  &= -\nabla p
     + \nabla \cdot \bigl(\nu_{\mathrm{eff}}\,\nabla\mathbf{U}\bigr),
 % \label{eq:mom_app}
  \end{aligned}
\end{equation}
where $\mathbf{U}=(u,v,w)$ is the Reynolds-averaged velocity, $p$ is the kinematic pressure, and
$\nu_{\mathrm{eff}}=\nu+\nu_t$ is the effective viscosity composed of the molecular viscosity $\nu$ and the eddy viscosity $\nu_t$.

The system is solved with OpenFOAM using the \texttt{simpleFoam} steady-state incompressible solver and the SIMPLE algorithm for pressure-velocity coupling. All simulations assume a neutrally stratified atmospheric boundary layer (ABL). Accordingly, no temperature (or potential-temperature) transport equation is solved and buoyancy does not contribute to the mean momentum balance, corresponding to the neutral atmospheric stability assumption standard for wind resource assessment~\cite{dhunny2017wind}.

% ---------------------------------------------------------
\subsection{Turbulence closure: $k$--$l$ model}

Turbulence is modeled using the $k$--$l$ framework originally proposed for geophysical flows by Mellor and Yamada~\cite{mellor1982development}. In the $k$--$l$ model, the turbulent kinetic energy $k$ is governed by a production--dissipation--diffusion balance. Under neutral stratification ($\partial\theta/\partial z=0$), the buoyancy production term vanishes and the transport equation reduces to:
\begin{equation}
  \frac{\mathrm{d}k}{\mathrm{d}t}
  = K_\text{M}\!\left[\left(\frac{\partial u}{\partial z}\right)^{\!2}
              +\left(\frac{\partial v}{\partial z}\right)^{\!2}\right]
    - \frac{(2k)^{3/2}}{B_1\,l}
    + \frac{\partial}{\partial z}\!\left(K_\text{E}\,\frac{\partial k}{\partial z}\right)
  \label{eq:k_transport}
\end{equation}
where $K_\text{M}$ and $K_\text{E}$ are the eddy diffusivities for momentum and turbulent kinetic energy, respectively, and $B_1$ is an empirical dissipation parameter.

\subsection{Calculation Domain}

As illustrated in Fig.~\ref{fig:dataset}b, the computational domain is defined in a three-dimensional Cartesian coordinate system, where the $x$-axis represents the longitudinal (west–east) direction, the $y$-axis represents the latitudinal (south–north) direction, and the $z$-axis denotes altitude. The entire domain is divided into three subregions: the Inner mountainous terrain, the Transition region, and the Coarse mesh region.

The Inner mountainous terrain corresponds to the real topographic area, characterized by $(\mathrm{dimensionX},$ $ \mathrm{dimensionY})$, which captures the major terrain variations. 
To ensure sufficient development of terrain-induced perturbations within a finite computational range, the outer layers are constructed as transition and buffer zones to smooth the flow field and maintain stable boundary conditions.

The radius of the circular computational domain is defined as:
\begin{equation}
\mathrm{radius} = 0.6\sqrt{2(\mathrm{dimensionX}^2 + \mathrm{dimensionY}^2)} + A
\label{eq:radius}
\end{equation}

In the vertical direction, the height of the domain is specified as:
\begin{equation}
\mathrm{dimensionZ} = \max(3.0\times \Delta_{\mathrm{MAP}},\, B)
\label{eq:height}
\end{equation}
where $\Delta_{\mathrm{MAP}}$ denotes the maximum elevation difference within the Inner mountainous terrain. The thickness of the Coarse mesh region is further defined as:
\begin{equation}
d_{\mathrm{max}} = 0.14\sqrt{\mathrm{dimensionX}^2 + \mathrm{dimensionY}^2},
\label{eq:dmax}
\end{equation}

The base elevation of the outer boundary is set to the mean elevation $\mathrm{avgZ}$ of the Inner region, ensuring both geometric and physical continuity between terrain and flow fields. Terrain surfaces are reconstructed from digital elevation data (SRTM at 30\,m resolution) and exported as triangulated surface (STL) files. Standard geometry checks are performed prior to meshing. A three-dimensional physical model is constructed from the terrain surface. The calculation Domain is predominantly hexahedral meshed, with spatially varying refinement in regions of strong topographic gradients (ridgelines, escarpments, valley flanks).

\subsection{Boundary conditions}

The cylindrical side surface is treated as a continuous boundary, as depicted in Fig.~\ref{fig:dataset}c. For each wind-direction sector, annulus segments where air enters the domain are assigned inlet conditions, while segments where air exits are treated as pressure outlets. On inflow segments, inlet profiles follow the neutral-ABL formulation of Richards and Hoxey~\cite{richards1993appropriate}:
\begin{align}
  U_{\mathrm{mag}}(z)
    &= \frac{u_\tau}{\kappa}
       \ln\!\left(\frac{z + z_0}{z_0}\right),
  \label{eq:loglaw_app}\\
  k(z)
    &= \frac{u_\tau^2}{\sqrt{C_\mu}}, \qquad C_\mu = 0.09,
  \label{eq:k_inlet}
\end{align}
where $u_\tau$ is the friction velocity, $\kappa=0.41$ is the von K\'arm\'an constant, and $z_0$ is the aerodynamic roughness length. In dataset generation, the reference wind speed is fixed to $U_{\text{mag}}^\text{ref} = 10\,\mathrm{m\,s^{-1}}$ at $z_{\mathrm{ref}} = 10$\,m for all simulations, while $z_0$ is assigned according to the roughness classification of each terrain tile.

The terrain surface is an adiabatic no-slip wall for $\mathbf{U}$, with rough-wall functions for turbulent quantities consistent with the prescribed roughness length $z_0$. The top face of the cylinder is treated as a symmetry boundary, minimizing artificial vertical confinement of the ABL flow. On outflow segments of the annulus, a pressure-outlet condition is prescribed (zero-gauge pressure) with zero-gradient conditions for velocity and turbulence variables to allow natural outflow.

To sample the full directional wind rose, each terrain tile is simulated under $N_\theta = 16$ uniformly spaced inflow angles:
\begin{equation}
  \theta_i = (i-1)\times 22.5^\circ, \quad i = 1, 2, \ldots, 16,
  \label{eq:angles}
\end{equation}

Rather than re-meshing for each direction, the cylindrical domain is kept fixed and the wind direction is changed by reassigning boundary condition types on the cylindrical side surface.

\section{Detailed dataset description}
\label{Detailed dataset description}

As summarized in Table~\ref{tab:core-crop-compare}, the raw dataset comprises 45 terrain types, 16 wind directions, and 467 samples, with each sample containing an average of approximately 0.55 million spatial points and 3.6 million edges, providing rich geometric and physical diversity for model learning.

\begin{table*}[ht!]
  \centering
  \caption{Comparison before and after core-region cropping. Reduction is $(\text{Raw}-\text{Processed})/\text{Raw}\times 100\%$.}
  \label{tab:core-crop-compare}
  \setlength{\tabcolsep}{6pt}
  \renewcommand{\arraystretch}{1.05}
  \begin{tabular}{l l c c c}
    \toprule
    \multirow{2}{*}{\textbf{Group}} & \multirow{2}{*}{\textbf{Metric}} & \multicolumn{3}{c}{\textbf{Statistics}} \\
    \cmidrule(lr){3-5}
      &  & \textbf{Raw} & \textbf{Processed} & \textbf{Reduction (\%)} \\
    \midrule
    \multirow{3}{*}{\textbf{Dataset}}
      & Samples  & 467 & 467 & --- \\
      & Terrains & 45  & 45  & --- \\
      & Angles   & 16  & 16  & --- \\
    \midrule
    \multirow{3}{*}{\textbf{Nodes}}
      & Count range   & 398345--766928 & 216046--369218 & --- \\
      & Count mean    & 551509.11      & 309635.99      & 43.9 \\
      & Count std. dev. & 47705.28     & 23161.29       & 51.4 \\
    \midrule
    \multirow{3}{*}{\textbf{Edges}}
      & Count range   & 2571808--5059320 & 1457592--2600408 & --- \\
      & Count mean    & 3603052           & 2116572.13       & 41.2 \\
      & Count std. dev. & 322420          & 173246.93        & 46.2 \\
    \midrule
    \multirow{3}{*}{\textbf{Grid size}}
      & Range         & 0--1063.01 & 0--285.72 & --- \\
      & Mean          & 138.42     & 43.00     & 68.9 \\
      & Std. dev.     & 148.73     & 32.11     & 78.4 \\
    \midrule
    \multirow{3}{*}{\textbf{Velocity} ($U_{\text{mag}}$)}
      & Range         & 0--26.17   & 0--26.17 & --- \\
      & Mean          & 12.73      & 10.62    & 16.6 \\
      & Std. dev.     & 6.66       & 5.77     & 13.4 \\
    \bottomrule
  \end{tabular}
\end{table*}

\section{Terrain attributes evaluation}
\label{terrain attr}

To quantitatively characterize the geometric complexity of mountainous terrain and establish a meaningful terrain classification framework for our dataset, we compute four key topographic metrics that collectively capture different aspects of terrain roughness and morphological variation. These metrics serve dual purposes: (1) they provide physically interpretable descriptors for analyzing terrain-flow coupling mechanisms, and (2) they enable quantitative assessment of whether our neural operators successfully generalize across the full spectrum of terrain complexities represented in the training distribution. Each metric is computed on a per-grid-point basis from the digital elevation model (DEM), yielding spatially resolved terrain characterization maps.

The local terrain slope $\alpha$ quantifies the maximum rate of elevation change at each grid point and is computed as:
\begin{equation}
\begin{aligned}
\alpha = \arctan\left(\sqrt{p^2 + q^2}\right)
\end{aligned}
\end{equation}
where $p$ and $q$ represent the terrain gradients in the $x$ and $y$ directions, respectively, evaluated using Horn's method~\cite{horn2005hill}:
\begin{equation}
\begin{aligned}
p_{\text{Horn}} &= \frac{(h_{++} + 2h_{+0} + h_{+-}) - (h_{-+} + 2h_{-0} + h_{--})}{8\Delta x}, \\
q_{\text{Horn}} &= \frac{(h_{++} + 2h_{0+} + h_{-+}) - (h_{+-} + 2h_{0-} + h_{--})}{8\Delta y}
\end{aligned}
\end{equation}
where $h_{ij}$ denotes the elevation at the neighbor located at relative position $(i,j)$ with respect to the central grid point, where $i,j \in \{-,0,+\}$ represent the left/center/right and bottom/center/top positions in the 3×3 neighborhood, and $\Delta x$, $\Delta y$ are the grid spacings. This weighted finite-difference scheme provides robust gradient estimation that is less sensitive to local elevation noise than simple two-point differences.

Following Dartnell~\cite{dartnell2000applying}, we define the roughness $R_i$ as the maximum elevation difference within the immediate neighborhood of grid point $i$:
\begin{equation}
\begin{aligned}
R_i = \max_{k \in \mathcal{N}(i)} |h_i - h_k|,
\label{eq:roughness}
\end{aligned}
\end{equation}
where $\mathcal{N}(i)$ denotes the eight-connected neighbors of point $i$. This metric captures abrupt topographic discontinuities such as cliffs, ridge crests, and valley bottoms that induce flow separation and strong velocity gradients. The unit is meters (or the unit of the DEM).

The rugosity coefficient $r_{\text{J}}$ quantifies three-dimensional surface complexity as the ratio of actual surface area to the corresponding planar projection, following Jenness~\cite{jenness2004calculating}:
\begin{equation}
\begin{aligned}
r_{\text{J}} = \frac{\sum_{k=1}^{8} A(T_{i,k})}{\Delta x \Delta y}
\end{aligned}
\end{equation}
where $A(T_{i,k})$ represents the three-dimensional area of the $k$-th triangular facet formed by connecting the central point $i$ to two adjacent neighbors in the 3×3 window, and $\Delta x \Delta y$ is the flat projected area. Values of $r_{\text{J}} \approx 1$ indicate nearly flat terrain, while $r_{\text{J}} > 1$ quantifies surface irregularity. This unitless area-based metric provides a complementary measure to roughness $R_i$ by accounting for cumulative terrain variations rather than just the maximum deviation.

 The TRI quantifies the root-mean-square elevation variability in the local neighborhood, as defined by Riley et al.~\cite{riley1999index}:
\begin{equation}
\begin{aligned}
\text{TRI}_i = \sqrt{\frac{1}{|\mathcal{N}(i)|}\sum_{k \in \mathcal{N}(i)} (h_i - h_k)^2},
\label{eq:tri}
\end{aligned}
\end{equation}
where $|\mathcal{N}(i)|$ denotes the cardinality of the neighborhood set (typically 8 for interior points). Unlike roughness, which captures only the maximum elevation span, TRI provides a statistically robust measure of overall terrain variability that correlates strongly with flow complexity and prediction difficulty in our experiments. The unit is meters.

These four complementary metrics, including slope (gradient magnitude), roughness (maximum elevation difference), rugosity (surface area ratio), and TRI (RMS elevation variability), collectively characterize terrain complexity from multiple geometric perspectives. During zero-shot evaluation on unseen real-world sites, we verify that the target terrains fall within the envelope defined by the training-set distributions of these attributes, thereby ensuring that generalization claims are not based on extrapolation beyond the model's experience.

\section{Zero-shot resolution robustness evaluation}
\label{appen: Zero-shot super-resolution analysis}

The core framework proposed in this work is designed to be resolution-robust, which is one of the major advantages of neural operator methods over conventional models defined on regular grids, such as convolution-based networks~\cite{ronneberger2015u} and Vision Transformers~\cite{dosovitskiy2020image}. In particular, our framework directly operates on unstructured point clouds and learns a continuous operator mapping from geometric-aware spatial samples to the corresponding wind field, rather than relying on a fixed lattice resolution. Therefore, after training, the model should ideally maintain stable predictive performance even when the number of query points observed in a single forward pass changes. This property is especially important for practical deployment, where the available point density may vary across computational budgets, domain discretizations, and hardware constraints.

To provide a strict and spatially complete evaluation, we adopt a partition-and-merge protocol in the zero-shot setting. For each full test point cloud, we divide the entire set of query points into disjoint subsets with inference ratios of 10\%, 20\%, 30\%, 40\%, 50\%, 70\%, 90\%, and 100\%. Each subset is independently fed into the trained model, and the predicted velocities are then restored to their original point indices and merged to reconstruct the full-field solution over the complete domain. For the last batch, the remaining query points are directly fed into the trained model. For example, when the ratio is set to 70\%, the model first processes 70\% of the points and then processes the remaining 30\%. This evaluation protocol keeps the terrain geometry, the target flow field, and the final evaluation coordinates unchanged, while varying only the number of points processed in each forward pass. This property is particularly important for wind field prediction over complex terrain, where the prediction error is often altitude-dependent. Therefore, performing inference on partitioned subsets and then merging the predictions back to the full domain provides a more reliable assessment of resolution robustness than directly evaluating on a randomly sparsified point cloud.

\begin{table*}[!htbp]
  \centering
  \caption{Performance comparison under different inference ratios for zero-shot resolution robustness evaluation}
  \label{tab:resolution-robustness}
  \footnotesize
  \setlength{\tabcolsep}{3.2pt}
  \renewcommand{\arraystretch}{1}
  \begin{tabular}{c*{14}{c}}
    \toprule
    \multirow{3}{*}{\textbf{Ratio}} &
      \multicolumn{14}{c}{\textbf{Metrics}} \\
    \cmidrule(lr){2-15}
      & \multicolumn{4}{c}{\textbf{MSE}}
      & \multicolumn{4}{c}{\textbf{L2} ($\%$)}
      & \multicolumn{4}{c}{\textbf{MAE}}
      & \multicolumn{2}{c}{\textbf{GPU (GB)}} \\
    \cmidrule(lr){2-5}\cmidrule(lr){6-9}\cmidrule(lr){10-13}\cmidrule(lr){14-15}
      & $u$ & $v$ & $w$ & {$U_{\text{mag}}$}
      & $u$ & $v$ & $w$ & {$U_{\text{mag}}$}
      & $u$ & $v$ & $w$ & {$U_{\text{mag}}$}
      & Avg. & Max. \\
    \midrule
    10\%  & 0.338  & 1.304 & 0.900 & 1.230 & 56.791 & 9.430 & 100.912 & 9.109 & 0.345 & 0.699 & 0.648 & 0.691 & 0.307 & 0.445 \\
    20\%  & 0.303 & 1.186 & 0.900 & 1.120 & 53.160 & 8.968 & 100.908 & 8.665 & 0.325& 0.668 & 0.647 & 0.660 & 0.566 & 0.771 \\
    30\%  & 0.296 & 1.162 & 0.900 & 1.097 & 52.403 & 8.867 & 100.908 & 8.572 &  0.321 & 0.662 & 0.647 & 0.654 & 0.872 & 1.142 \\
    40\%  & 0.289 & 1.138 & 0.900 & 1.075 & 51.646 & 8.769 & 100.905 & 8.479 & 0.317 & 0.656 & 0.647 & 0.648 & 1.136 & 1.443  \\
    50\%  & 0.282 &  1.113 & 0.900 & 1.053 & 50.838 & 8.667 & 100.904 & 8.383 & 0.313 & 0.649 & 0.647 & 0.641 & 1.464 & 1.798 \\
    70\%  & 0.282 & 1.114 & 0.900 & 1.054 & 50.880 & 8.669 & 100.904 & 8.387 & 0.313 & 0.649 & 0.647 & 0.641 & 2.031 & 2.455 \\
    90\%  & 0.282 & 1.112 & 0.900 & 1.053 & 50.893 & 8.662 & 100.903 & 8.382 & 0.313 & 0.649 & 0.647 & 0.641 & 2.442 & 3.119 \\
    100\% & 0.275 & 1.089 & 0.900 & 1.032 & 50.100 & 8.564 & 100.902 & 8.290 & 0.309 & 0.642 & 0.647 & 0.635  & 2.938 & 3.675 \\
    \bottomrule
  \end{tabular}
  \vspace{2pt}
  \begin{minipage}{\textwidth}
  \footnotesize \textit{Note:} GPU denotes the GPU memory consumption during inference. The reported Avg.\ and Max.\ values are calculated over all test cases in the test set based on the peak memory usage of each case, and therefore depend on the number of points contained in the input point cloud.
  \end{minipage}
\end{table*}
% {\footnotesize \textit{Note:} GPU denotes the GPU memory consumption during inference. The reported Avg.\ and Max.\ values are calculated over all test cases in the test set based on the peak memory usage of each case, and therefore depend on the number of points contained in the input point cloud.}

The full zero-shot test results are summarized in Table~\ref{tab:resolution-robustness}, including both prediction errors and GPU memory usage during inference. Overall, the results show a clear trade-off between predictive accuracy and memory efficiency as the inference ratio (spatial resolution) decreases. It can be seen that lower inference ratios lead to consistently higher errors, while substantially reducing the GPU memory required in each forward pass. Specifically, for example, at an inference ratio of 10\%, the MSE and relative L2 error of $U_{\text{mag}}$ increase to 1.230 and 9.109\%, corresponding to relative increases of approximately 19\% and 9.9\% over the full-resolution setting (100\%). Meanwhile, the maximum GPU memory usage is reduced from 3.675~GB to 0.445~GB, which is close to 12\% of the full-resolution requirement. In addition, the error curves become nearly flat once the inference ratio reaches 50\% or above. The differences among the 50\%, 70\%, and 90\% settings are minimal for all reported metrics, indicating that the model has already captured most of the effective spatial information at these ratios. Taken together, these results demonstrate that the proposed framework achieves strong zero-shot resolution robustness, while also offering flexible control over the trade-off between accuracy and computational cost.

\section{Zero-shot performance comparison across unseen mountain sites}
\label{appen:Zero-shot performance comparison }

The detailed inference results across unseen mountain sites are summarized in Table~\ref{tab:zero-shot-all}.

\begin{table*}[ht!]
  \centering
  \caption{Zero-shot performance comparison across unseen mountain sites (1: Chatou-1, 2: Chatou-2, 3: Daguping, 4: Hengdong).}
  \label{tab:zero-shot-all}
  \setlength{\tabcolsep}{2pt}
  \renewcommand{\arraystretch}{1}
  \begin{tabular}{l l *{12}{c}}
    \toprule
    \multirow{3}{*}{\textbf{Site}} & \multirow{3}{*}{\textbf{Model}} & \multicolumn{12}{c}{\textbf{Metrics}} \\
    \cmidrule(lr){3-14}
      & & \multicolumn{4}{c}{\textbf{MSE}}
        & \multicolumn{4}{c}{\textbf{L2 (\%)}}
        & \multicolumn{4}{c}{\textbf{MAE}} \\
    \cmidrule(lr){3-6}\cmidrule(lr){7-10}\cmidrule(lr){11-14}
      & & $u$ & $v$ & $w$ & {$U_{\text{mag}}$}
        & $u$ & $v$ & $w$ & {$U_{\text{mag}}$}
        & $u$ & $v$ & $w$ & {$U_{\text{mag}}$} \\
    \midrule
    \multirow{4}{*}{1}
      & Transolver    & 0.204 & 1.137 & 1.003 & 1.120 & 42.731 & 8.753 & 100.583 & 8.617 & 0.267 & 0.700 & 0.689 & 0.705 \\
      & AeroGTO       & 0.405 & 1.745 & 0.115 & 1.681 & 68.256 & 10.626 & 39.207 & 10.375 & 0.379 & 0.911 & 0.227 & 0.901 \\
      & Patch-solver  & 0.198 & 1.164 & 0.990 & 0.828 & 41.838 & 8.679 & 99.989 & 7.408 & 0.265 & 0.726 & 0.692 & 0.580 \\
      & Patch-GTO     & 0.624 & 0.863 & 0.089 & 1.191 & 84.742 & 7.620 & 34.419 & 8.733 & 0.482 & 0.583 & 0.195 & 0.736 \\
    \midrule
    \multirow{4}{*}{2}
      & Transolver    & 0.181 & 0.892 & 0.688 & 0.849 & 50.294 & 7.518 & 100.586 & 7.299 & 0.241 & 0.549 & 0.572 & 0.545 \\
      & AeroGTO       & 0.288 & 0.892 & 0.083 & 0.863 & 69.998 & 7.536 & 34.238 & 7.383 & 0.324 & 0.563 & 0.199 & 0.558 \\
      & Patch-solver  & 0.145 & 0.693 & 0.688 & 0.662 & 45.156 & 6.622 & 100.617 & 6.444 & 0.227 & 0.499 & 0.580 & 0.496 \\
      & Patch-GTO     & 0.371 & 0.696 & 0.071 & 0.682 & 79.351 & 6.657 & 31.644 & 6.565 & 0.383 & 0.510 & 0.171 & 0.510 \\
    \midrule
    \multirow{4}{*}{3}
      & Transolver    & 0.500 & 2.093 & 1.540 & 1.889 & 51.009 & 13.615 & 100.511 & 12.739 & 0.448 & 0.929 & 0.884 & 0.895 \\
      & AeroGTO       & 0.748 & 1.676 & 0.219 & 1.519 & 80.646 & 11.698 & 50.651  & 11.048 & 0.569 & 0.844 & 0.330 & 0.808 \\
      & Patch-solver  & 0.461 & 1.683 & 1.527 & 1.515 & 49.015 & 12.184 & 100.202 & 11.391 & 0.443 & 0.857 & 0.884 & 0.829 \\
      & Patch-GTO     & 0.982 & 1.574 & 0.162 & 1.444 & 92.395 & 11.336 & 43.573  & 10.772 & 0.641 & 0.793 & 0.276 & 0.771 \\
    \midrule
    \multirow{4}{*}{4}
      & Transolver    & 0.066 & 0.342 & 0.157 & 0.336 & 68.470 & 4.537  & 100.642 & 4.491 & 0.149 & 0.339 & 0.251 & 0.337 \\
      & AeroGTO       & 0.100 & 0.297 & 0.035 & 0.284 & 87.612 & 4.209  & 49.252  & 4.111 & 0.191 & 0.311 & 0.131 & 0.309 \\
      & Patch-solver  & 0.044 & 0.240 & 0.176 & 0.238 & 55.927 & 3.800  & 106.524 & 3.776 & 0.132 & 0.287 & 0.286 & 0.286 \\
      & Patch-GTO     & 0.106 & 0.275 & 0.029 & 0.262 & 89.799 & 4.051  & 44.558  & 3.952 & 0.198 & 0.266 & 0.105 & 0.264 \\
    \bottomrule
  \end{tabular}
\end{table*}

\section{Training configuration}
\label{traing config}

\paragraph{\textbf{Baselines}}
We comprehensively compare our Patch-solver and Patch-GTO against state-of-the-art (SOTA) DL methods for PDE solving, covering diverse architectural paradigms including Transformer-based, Fourier Neural Operator (FNO)-based, and graph-based approaches:

\begin{enumerate}
    \item \textbf{Transolver}~\cite{wu2024transolver}: a Transformer-based universal PDE solver with global physics attention mechanism;
    \item \textbf{AeroGTO}~\cite{liu2025aerogto}: a graph Transformer operator specifically designed for aerodynamic flow predictions;
    \item \textbf{GINO}~\cite{li2023geometry}: a geometry-informed neural operator that first maps irregular geometries into a latent regular representation via GNO \cite{anandkumar2020neural} layers and then applies an FNO backbone for operator learning;
    \item \textbf{GNOT}~\cite{hao2023gnot}: a graph neural operator that leverages Transformer-based message aggregation for PDE solving;
    \item \textbf{Geo-FNO}~\cite{li2023fourier}: a geometry-aware neural operator that incorporates geometry through coordinate-aware inputs and applies an FNO backbone on a regular reference grid.
\end{enumerate}

It is important to note that our Patch-solver and Patch-GTO are built upon the core architectures of Transolver and AeroGTO, respectively, by incorporating our proposed dual-attention mechanism. Specifically, Patch-solver extends Transolver's global slice attention with an additional local sectional attention to better capture terrain-induced flow features, while Patch-GTO integrates the same dual-attention strategy into AeroGTO's graph Transformer framework. This design enables direct comparative analysis to validate the effectiveness of our dual-attention mechanism for complex terrain wind field prediction. 

For all baseline models, we adopt their recommended hyperparameter configurations from the official code repositories and train them on our dataset to ensure fair comparison. All models are trained for 100 epochs under identical experimental settings, including the same batch size (1) and evaluation protocol on the same test set.

\paragraph{\textbf{Input formation and normalization}} 
For Patch-solver, each query point is represented by a 7D feature vector
$[x,y,z,\,\hat n_x,\hat n_y,\hat n_z,\,d]$, where $(x,y,z)$
are the Cartesian coordinates, $\hat{\boldsymbol n}=\nabla d/\|\nabla d\|$
is the unit normal recovered from the signed distance field (SDF), and $d$ is the signed distance from the query point to the nearest terrain surface (positive in the fluid domain above the terrain).

% For Patch-GTO, we construct a graph $\mathbf{G}=(\mathbf{V},\mathbf{E})$ on the same points. Each node $\mathbf{V}_i$ carries the \emph{same} 3D feature $\mathbf v_i=[x_i,y_i,z_i]$, while edges ($\mathbf{E}_{ij}$) are built with a fixed $k$-NN neighborhood in coordinate space. For every edge $\mathbf{E}_{ij}$, we use the relative-position attribute
% $\mathbf E_{ij}=[\,\mathbf p_j-\mathbf p_i,\,\|\mathbf p_j-\mathbf p_i\|_2\,]$ with $\mathbf p_i=[x_i,y_i,z_i]$; this supplies Patch-GTO with both directional and metric cues.

For Patch-GTO, we construct a graph $\mathbf{G}=(\mathbf{V},\mathbf{E})$ on the same points. Each node $\mathbf{V}_i$ carries the same 3D coordinate feature $\mathbf{v}_i = [x_i, y_i, z_i]$. The edge set $\mathbf{E} \in \mathbb{Z}^{N_\text{e} \times 2}$ is constructed by combining the original mesh connectivity with additional edges established via $k$-NN search in coordinate space. For each edge $(i,j) \in \mathbf{E}$, we compute the relative-position attribute $\mathbf{e}_{ij} = [\mathbf{p}_j - \mathbf{p}_i,\, \|\mathbf{p}_j - \mathbf{p}_i\|_2]$ with $\mathbf{p}_i = [x_i, y_i, z_i]$; this supplies Patch-GTO with both directional and metric cues.

All feature channels and velocity targets are standardized to zero mean and unit variance using statistics from the training split. During inference, we undo the normalization to report errors in physical units.

\paragraph{\textbf{Hyperparameter setting}}
For the proposed Patchsolver and Patch-GTO, the PCPM partitions the domain into $(n_x,n_y,n_z)=(10,10,10)$ voxels unless specified otherwise, and we keep the inverse permutation to map predictions back to the original mesh order. The PEM uses $H\!=\!164$ hidden size, $L\!=\!4$ PDAM blocks, $H_{\!n}\!=\!8$ attention heads, $4H$ feed-forward expansion. Slice-based global attention compresses the domain to $G\!=\!16$ tokens per head; local sectional attention is applied within each voxel group. The fusion gate in Eq.~(\ref{eq:FINAL}) is initialized so that $\sigma(g)\!=\!0.5$.

\paragraph{\textbf{Optimization and Hardware}}

All models are trained with Adam (weight decay $10^{-4}$, $\beta_1\!=\!0.9$, $\beta_2\!=\!0.999$) under a OneCycleLR schedule; the initial learning rate is $1\!\times\!10^{-3}$. All the experiments are conducted on a single NVIDIA A800 with a batch size of 1 and total training epochs of 100. Inference is single-pass and scales linearly with the number of query points. For cases in this paper (around $3\times10^6$ mesh points for single case), Patch-solver achieves an average inference time of 0.73~s per case, corresponding to a throughput of $\sim 4.1{\times}10^{6}$ points/s.

\paragraph{\textbf{Evaluation metrics}}
To clearly distinguish the predicted results, the velocity components, $u$, $v$, $w$, and the wind speed magnitude $U_\text{mag}$ are evaluated separately. The error metrics, MSE, MAE, and relative L2 loss are used to measure the prediction performance for all algorithms. The relative L2 error reports the overall Euclidean misalignment of the predicted and true velocity fields, and is normalized by the ground-truth vector so that the result is dimensionless and comparable across datasets, calculated by Eq. \ref{eq1: l2}. MSE loss (Eq. \ref{eq1: mse}) calculates the average squared deviations, and MAE (Eq. \ref{eq1: mae}) measures the average absolute deviations.
\begin{equation}
\mathrm{L2}
  \;=\;
  \frac{\lVert\,\widehat{\mathbf{Y}}-\mathbf{Y}\,\rVert_{2}}
       {\lVert\,\mathbf{Y}\,\rVert_{2}}
  \;=\;
  \frac{\displaystyle
        \sqrt{\sum_{i=1}^{n}\bigl(\hat {\mathbf{y}}_i - \mathbf{y}_i\bigr)^2}}
       {\displaystyle
        \sqrt{\sum_{i=1}^{n}\mathbf{y}^{\,2}}}.
\label{eq1: l2}
\end{equation}

\begin{equation}
  \mathrm{MSE}
  =\frac{1}{N}\sum_{i=1}^{N}\bigl(\hat {\mathbf{y}}_i-\mathbf{y}_i\bigr)^2
\label{eq1: mse}
\end{equation}
\begin{equation}
  \mathrm{MAE}
  =\frac{1}{N}\sum_{i=1}^{N}\bigl|\hat {\mathbf{y}}_i-\mathbf{y}_i\bigr|
\label{eq1: mae}
\end{equation}
where $\hat{\mathbf y}$ and ${\mathbf y}$ are the predicted and the ground-truth vectors, respectively. $N$ is the number of points in vector ${\mathbf y}$.

\paragraph{\textbf{Computational cost}}

For CFD data generation, each OpenFOAM simulation requires approximately 1 CPU-core-hour on an AMD EPYC 7543 processor. In total, the full dataset generation requires about 580 CPU-core-hours for 530 cases (467 training/test cases plus 63 zero-shot cases). For the surrogate model, Patch-solver performs inference in approximately 0.73~s per case on a single NVIDIA A800 GPU for around 300,000 mesh points. During training, Patch-solver requires 143 GPU-hours for 100 epochs on one NVIDIA A800 GPU, with a peak GPU memory usage of 55~GiB.

\paragraph{\textbf{Training process}}

\begin{figure}[htbp!]
\centering
\includegraphics[width=\textwidth]{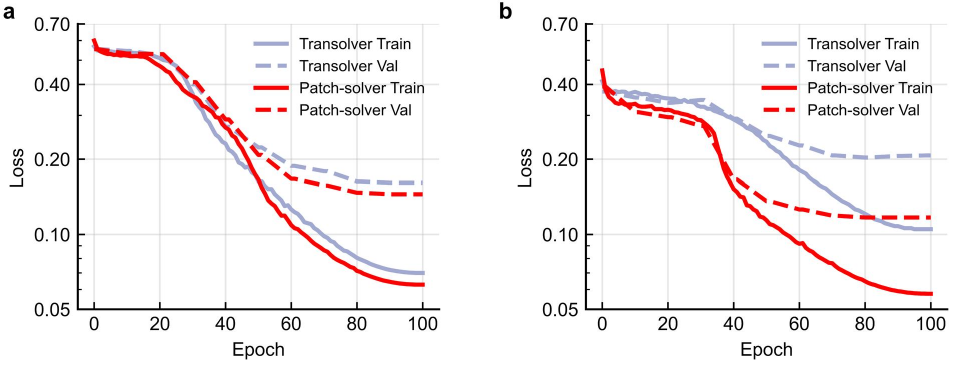}
\caption{\textbf{Training processes of Transolver and Patch-solver under different input settings.} \textbf{a} Training and validation loss curves for the scenario without sparse sensor input. \textbf{b} Training and validation loss curves for the scenario with sparse sensor input.}
\label{fig:loss-epoch}
\end{figure}

Fig.~\ref{fig:loss-epoch} shows the evolution of training and validation losses for Transolver and Patch-solver in two settings: without sparse sensor input and with sparse sensor input. In the setting without sparse observations, both models reach a stable regime at around 80 epochs, while Patch-solver generally maintains slightly lower training and validation losses throughout the optimization process. When sparse sensor data are incorporated as additional input, the convergence of Patch-solver is further accelerated, reaching a stable plateau at around 70 epochs. More importantly, the gap between Patch-solver and Transolver becomes more pronounced in this sparse-data setting, suggesting that the proposed dual-attention design is more effective at assimilating sparse observational constraints.

\section{Data size analysis}
\label{Data size analysis}

To further examine how the amount and distribution of training data affect zero-shot generalization, we conduct a training-set size sensitivity analysis using the proposed Patch-solver model. The baseline dataset used in the main text consists of 45 terrain geometries and 467 terrain-angle combinations generated from CFD simulations. In this section, we construct several training subsets with different sizes, namely 36\%, 68\%, 100\%, and 135\%, while keeping the zero-shot evaluation protocol unchanged. Here, 100\% denotes the formal training dataset adopted in the main text, and 135\% denotes an expanded dataset obtained by adding extra terrain-flow samples beyond this baseline. Table~\ref{tab:zero-shot-datasetana} summarizes the zero-shot performance on four unseen mountainous sites with various inflow angles. Overall, increasing the training-set size from 36\% to 68\% and then to 100\% leads to substantial and consistent improvements across all sites, especially for the primary flow direction $v$. These results indicate that broader terrain-flow exposure during training significantly improves the transferability of the learned operator to geographically unseen sites.

\begin{table*}[ht!]
  \centering
  \caption{Zero-shot performance comparison across unseen mountain sites with various training set sizes (1: Chatou-1, 2: Chatou-2, 3: Daguping, 4: Hengdong). The Patch-solver model architecture is selected for zero-shot evaluation. The 100\% in the table denotes the training dataset kept as the formal analysis in the main text.}
  \label{tab:zero-shot-datasetana}
  \setlength{\tabcolsep}{2pt}
  \renewcommand{\arraystretch}{1}
  \begin{tabular}{l l *{12}{c}}
    \toprule
    \multirow{3}{*}{\textbf{Site}} & \multirow{3}{*}{\textbf{ Data size}} & \multicolumn{12}{c}{\textbf{Metrics}} \\
    \cmidrule(lr){3-14}
      & & \multicolumn{4}{c}{\textbf{MSE}}
        & \multicolumn{4}{c}{\textbf{L2 (\%)}}
        & \multicolumn{4}{c}{\textbf{MAE}} \\
    \cmidrule(lr){3-6}\cmidrule(lr){7-10}\cmidrule(lr){11-14}
      & & $u$ & $v$ & $w$ & {$U_{\text{mag}}$}
        & $u$ & $v$ & $w$ & {$U_{\text{mag}}$}
        & $u$ & $v$ & $w$ & {$U_{\text{mag}}$} \\
    \midrule
    \multirow{4}{*}{1}
      & 36\%    & 0.942 & 3.002 & 0.996 & 2.709 &   91.885& 14.181 & 100.927 & 13.287  & 0.634 & 1.149 & 0.688 & 1.114 \\
      & 68\%       & 0.584  & 2.031 & 0.978 &  1.917 & 72.125 & 11.673 & 100.049 & 11.248 & 0.459 & 0.974 & 0.690  & 0.964 \\
      &  100\%  & 0.198 & 1.164 & 0.990 & 0.828 & 41.838 & 8.679 & 99.989 & 7.408 & 0.265 & 0.726 & 0.692 & 0.580 \\
      & 135\%    & 0.246 & 1.216  & 0.976 &  1.197 & 46.863 & 9.022  & 100.004 &  8.889 & 0.296 &  0.739 & 0.684  & 0.741 \\
    \midrule
    \multirow{4}{*}{2}
      & 36\%    & 0.631 & 2.414 & 0.706 & 2.309 &  93.962 & 12.358 & 101.805 & 12.028 & 0.500 & 1.051  & 0.580 & 1.044\\
      & 68\%      &  0.406 & 1.601 & 0.678 & 1.460 & 75.124 & 10.072 & 99.706 & 9.573 & 0.385 & 0.833 & 0.576 & 0.817 \\
      &  100\%  & 0.145 & 0.693 & 0.688 & 0.662 & 45.156 & 6.622 & 100.617 & 6.444 & 0.227 & 0.499 & 0.580 & 0.496 \\
      & 135\%    & 0.180 & 0.898 & 0.681 & 0.845 & 50.207 & 7.542 & 100.007 & 7.286 & 0.257 & 0.590 & 0.573  & 0.581\\
    \midrule
    \multirow{4}{*}{3}
      & 36\%    & 2.316 & 5.850 & 1.517 & 4.647 & 105.683 & 22.649 & 101.038 & 19.901 & 1.038 & 1.671 & 0.878 & 1.499 \\
      & 68\%       & 1.182 & 3.773 & 1.472 & 3.046 & 74.902 & 18.223 & 99.481 & 16.138 & 0.405 & 1.325 & 0.869  & 1.207 \\
      &  100\%  & 0.461 & 1.683 & 1.527 & 1.515 & 49.015 & 12.184 & 100.202 & 11.391 & 0.443 & 0.857 & 0.884 & 0.829 \\
      & 135\%     & 0.561 & 2.054 & 1.480 & 1.878 & 53.675 & 13.451 & 99.750 & 12.661 & 0.489 & 0.974 & 0.869 & 0.941\\
    \midrule
    \multirow{4}{*}{4}
      & 36\%     & 0.170  & 0.660 & 0.171 & 0.656 & 109.800 & 6.290 & 105.073 & 6.265 & 0.273 & 0.518 & 0.273  & 0.519 \\
      & 68\%        & 0.099 & 0.514 & 0.166 & 0.507 & 83.795 & 5.559  & 103.666 & 5.516 & 0.193 & 0.490 & 0.264  & 0.489 \\
      &  100\%  & 0.044 & 0.240 & 0.176 & 0.238 & 55.927 & 3.800  & 106.524 & 3.776 & 0.132 & 0.287 & 0.286 & 0.286 \\
      & 135\%    &  0.042 & 0.285 & 0.160 & 0.281 & 54.559 & 4.135 & 101.830 & 4.105 & 0.124 & 0.317 & 0.261 & 0.316\\
    \bottomrule
  \end{tabular}
\end{table*}

\begin{figure}[htbp!]
\centering
\includegraphics[]{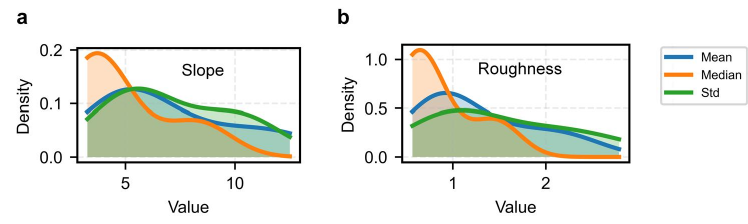}
\caption{\textbf{Kernel–density distributions of two terrain descriptors (Slope and Roughness) computed over the newly added data.} \textbf{a} Slope \textbf{b} Roughness}
\label{fig:newly added data}
\end{figure}

However, the performance gain is not strictly monotonic when the training-set size is further increased to 135\%. Compared with the 100\% setting, the $U_{\text{mag}}$ relative $L_2$ error slightly increases. This result suggests that zero-shot generalization is not determined solely by the number of training samples. Instead, it also depends on whether the newly added samples effectively enlarge the terrain-statistics support represented in the training distribution. To better understand this behavior, Fig.~\ref{fig:newly added data} presents the kernel-density distributions for the newly added samples. Compared with the original training set in the main text (Fig. \ref{fig:result1}), the newly added terrains are concentrated mainly in the lower-complexity regime, with distributions biased toward relatively smaller slope and roughness values.  As a result, the added cases introduce redundancy in already well-represented regions of the terrain space, while contributing limited new information for the unseen, more challenging zero-shot sites.

Overall, the results in Table~\ref{tab:zero-shot-datasetana} and Fig.~\ref{fig:newly added data} support a more precise interpretation of generalization in the present work: the proposed framework exhibits zero-shot transfer within the support of the terrain statistics covered by the training dataset, while its performance remains sensitive to distributional imbalance in the added training samples. Importantly, future dataset expansion should focus more on terrains with higher complexity compared with the training dataset, thereby broadening the effective support of the training distribution and improving the model’s generalization performance on unseen mountainous sites.

\end{appendices}





% \end{thebibliography}

% \begin{appendices}

% \end{appendices}

\end{linenumbers}

\end{document}